\documentclass{article}
\usepackage{bussproofs}
\EnableBpAbbreviations
\usepackage{xspace}
\usepackage{amssymb}
\usepackage{amsmath}
\usepackage{latexsym}
\usepackage{url}
\usepackage{amsthm}
\theoremstyle{plain}
\newtheorem{Th}{Theorem}[section]
\newtheorem{prop}[Th]{Proposition}
\newtheorem{lem}[Th]{Lemma}
\newtheorem{cor}[Th]{Corollary}
\newtheorem{observation}[Th]{Observation}

\theoremstyle{definition}
\newtheorem{DEF}[Th]{Definition}

\newtheorem{REM}[Th]{Remark}

\newtheorem{notation}[Th]{Notation}

\def\NN{\mathbb{N}} 
\def\OO#1{\mathcal O({#1})}

\def\D{\mathcal D}
\def\sz#1{|{#1}|}
\def\oo#1{o({#1})}

\def\I{\mathsf I}
\def\R{\mathsf R}

\def\E{\mathsf E}

\def\Ecl#1{\widetilde{#1}}

\def\rstr#1#2{{#1}_{#2}}

\def\Rule#1#2{\AxiomC{$#1$}\UnaryInfC{$#2$}\DisplayProof}
\def\RuleN#1{\AxiomC{\strut}\UnaryInfC{$#1$}\DisplayProof}
\def\RuleB#1#2#3{\AxiomC{$#1$}\AxiomC{$#2$}\BinaryInfC{$#3$}\DisplayProof}

\def\Sz{\mathfrak S}
\def\szf#1{\vartheta({#1})}
\def\szfe#1#2{\vartheta_{#1}({#2})}
\def\szfk#1{\szfe{k}{#1}}

\def\varDeriv#1#2{{#1}\colon{#2}}

\newcommand{\krajicek}{Kraj{\'\i\v{c}}ek\xspace}

\newcommand{\setof}[2]{\left\{ #1 \colon #2 \right\}}
\newcommand{\seq}[1]{\left\langle #1 \right\rangle}

\newcommand{\pow}{\mathfrak{P}}
\newcommand{\finpow}{\pow_\text{fin}}
\newcommand{\finpot}{\finpow}

\newcommand{\verum}{\top}
\newcommand{\falsum}{\bot}
\newcommand{\tbv}{{\textstyle\bigvee}}           
\newcommand{\tbw}{{\textstyle\bigwedge}}        
\newcommand{\und}{\mathrel{\:\land\:}}
\newcommand{\oder}{\mathrel{\:\lor\:}}
\newcommand{\imp}{\mathrel{\:\rightarrow\:}}

\newcommand{\lba}{\ensuremath{\cL_{\mathrm{BA}}}\xspace}
\newcommand{\BFOR}{\ensuremath{\mathrm{BFOR}}\xspace}
\newcommand{\FOR}{\BFOR}
\DeclareMathOperator{\suc}{s}

\newcommand{\sma}{\mathbin{\#}}
\newcommand{\hilfminus}[1]{\stackrel{#1\cdot}{\relbar}}
\newcommand{\aminus}{\mathbin{\mathchoice {\hilfminus{\displaystyle}}%
        {\hilfminus{\textstyle}}{\hilfminus{\scriptstyle}}%
        {\hilfminus{\scriptscriptstyle}}}}
\newcommand{\hilfshr}[2]{{{#1\lfloor\frac{1}{2}} #2 {#1\rfloor}}}
\newcommand{\shr}[1]{\mathbin{\mathchoice {\hilfshr{\textstyle}{#1}}%
        {\hilfshr{\textstyle}{#1}}{\hilfshr{\scriptstyle}{#1}}%
        {\hilfshr{\scriptscriptstyle}{#1}}}}

\newcommand{\zhl}[1]{2^{| #1 |}}
\newcommand{\kl}{\mathbin{<}}
\newcommand{\klg}{\mathbin{\le}}

\newcommand{\ba}{\mathrm{BA}}
\newcommand{\sib}[1]{\ensuremath{\Si^{\mathrm{b}}_{#1}}\xspace}
\newcommand{\pib}[1]{\ensuremath{\Pi^{\mathrm{b}}_{#1}}\xspace}

\newcommand{\ssib}[1]{\ensuremath{\mathrm{s}\Si^{\mathrm{b}}_{#1}}\xspace}

\newcommand{\basic}{\ensuremath{\mathrm{BASIC}}\xspace}
\newcommand{\Ind} {\mathrm{Ind}}
\newcommand{\ind} {\mathrm{IND}}

\newcommand{\ilind}[1]{\mathrm{L}^{#1}\mathrm{IND}}

\newcommand{\dilind}[1]{\mbox{-}\ilind{#1}}

\newcommand{\rS}{\mathrm{S}}

\newcommand{\infConn}{\ensuremath{\mathbb{C}}\xspace}
\newcommand{\Finf}{\ensuremath{\mathcal{F}_\infty}\xspace}
\newcommand{\ieval}[1]{{[\![#1]\!]_\infty}}

\newcommand{\ForN}{\ensuremath{\mathcal{F}}\xspace}
\newcommand{\ForNBA}{\ensuremath{\mathcal{F}_\mathrm{BA}}\xspace}

\newcommand{\DerN}{\ensuremath{\mathcal{H}}\xspace}
\newcommand{\ceDerN}{\ensuremath{\mathcal{C}\mathcal{H}}\xspace}

\newcommand{\DerNBA}{\ensuremath{\mathcal{H}_\mathrm{BA}}\xspace}
\newcommand{\ceDerNBA}{\ensuremath{\mathcal{C}\mathcal{H}_\mathrm{BA}}\xspace}
\newcommand{\compDerNBA}{\ensuremath{\mathcal{C}\mathrm{omp}\mathcal{H}_\mathrm{BA}}\xspace}
\newcommand{\abstr}[1]{\overline{#1}}
\newcommand{\eval}[1]{{[\![#1]\!]}}

\newcommand{\ieq}{\approx}
\newcommand{\seqieq}[1]{{\ieq}#1}
\newcommand{\red}{\mathrel{\rightarrow^1_\NN}}
\newcommand{\eqnat}{\mathrel{\ieq_\NN}}
\newcommand{\seqeqnat}[1]{{\eqnat}#1}

\newcommand{\PrSys}{\mathfrak{S}}
\newcommand{\IS}{\mathcal{I}}
\DeclareMathOperator{\last}{last}
\DeclareMathOperator{\ord}{o}
\DeclareMathOperator{\tp}{tp}
\DeclareMathOperator{\crk}{crk}
\DeclareMathOperator{\hgt}{hgt}
\DeclareMathOperator{\dszop}{sz}
\newcommand{\dsz}[1]{\dszop(#1)}
\newcommand{\PrSysInf}{\PrSys_\infty}
\newcommand{\PrSysForN}{\PrSys_\ForN}
\newcommand{\bainf}{\ensuremath{\ba^\infty}\xspace}
\newcommand{\bastar}{\ensuremath{\ba^\star}\xspace}

\newcommand{\Ax}{\mathrm{Ax}}
\newcommand{\Cut}{\mathrm{Cut}}
\newcommand{\Rep}{\mathrm{Rep}}

\newcommand{\opI}{\mathbb{I}}
\newcommand{\opR}{\mathbb{R}}
\newcommand{\opE}{\mathbb{E}}
\newcommand{\derd}{d}

\DeclareMathOperator{\rk}{rk}

\DeclareMathOperator{\fv}{FV}
\DeclareMathOperator{\bd}{bd}
\DeclareMathOperator{\ibd}{ibd}
\DeclareMathOperator{\rng}{rng}
\DeclareMathOperator{\deco}{deco}
\DeclareMathOperator{\card}{card}
\newcommand{\ork}[1]{\mathop{#1\mbox{-\textup{rk}}}}
\newcommand{\Crk}{\ork{\cC}}
\newcommand{\ocrk}[1]{\mathop{#1\mbox{-\textup{crk}}}}
\newcommand{\Ccrk}{\ocrk{\cC}}

\newcommand{\Pt}{\mathrm{P}}
\newcommand{\FP}{\mathrm{FP}}
\newcommand{\wit}{\text{wit}}

\newcommand{\cC}{\mathcal{C}}

\newcommand{\cF}{\mathcal{F}}
\newcommand{\cI}{\mathcal{I}}
\newcommand{\cL}{\mathcal{L}}

\newcommand{\al}{\alpha}
\newcommand{\be}{\beta}

\newcommand{\om}{\omega}

\newcommand{\vhi}{\varphi}
\newcommand{\De}{\Delta}
\newcommand{\Ga}{\Gamma}
\newcommand{\Si}{\Sigma}

\newcommand{\Und}{\ensuremath{\quad\&\quad}}
\newcommand{\Imp}{\ensuremath{\quad\Rightarrow\quad}}
\newcommand{\Gdw}{\ensuremath{\quad\Leftrightarrow\quad}}

\newcommand{\noi}{\noindent}
\newcommand{\ul}{\underline}

\begin{document}

\title{On the computational complexity of cut-reduction}
\author{Klaus Aehlig\\
Computer Science\\
Swansea University\\
Swansea SA2 8PP, UK\\
\url{k.t.aehlig@swansea.ac.uk}
 \and Arnold Beckmann\\
Computer Science\\
Swansea University\\
Swansea SA2 8PP, UK\\
\url{a.beckmann@swansea.ac.uk}}

\maketitle

\begin{abstract}
We investigate the complexity of cut-reduction on
proof notations, in particular identifying situations where cut-reduction
operates feasibly, i.e., sub-exponential, on proof notations.
We then apply the machinery to characterise definable search problem in
Bounded Arithmetic.

To explain our results with an example, 
let $\opE(d)$ denote Mints' continuous cut-reduction operator which reduces 
the complexity of all cuts of a propositional derivation $d$ by one level.
We will show that if all sub-proofs of $d$ 
can be denoted with notations of size $s$, and the height of $d$ is $h$, then
sub-proofs of the derivation $\opE(d)$ 
can be denoted by notations of size $h\cdot(s+\OO1)$.
Together with the observation that determining the last inference of a denoted
derivation as well as determining notations for immediate sub-derivations is easy 
(i.e., polynomial time computable),
we can apply this result to re-obtain that the $\sib i$-definable 
functions of the Bounded Arithmetic theory $\rS^i_2$ are in the $i$-th level of the 
polynomial time hierarchy of functions $\FP^{\sib{i-1}}$.

\end{abstract}

\begin{section}{Introduction and Related Work}
\def\rep{\mathcal R}
Since Gentzen's invention of the ``Logik Kalk{\"u}l'' LK and the proof of his 
``Hauptsatz'' \cite{Gentzen35,Gentzen35a}, 
cut-elimination has been studied in many papers on proof theory.
Mints' invention of continuous normalisation \cite{Mints75,KreiselMintsSimpson75}
isolates operational aspects of normalisation, that is 
the manipulations on (infinitary) propositional derivations.
These operational aspects are described independently of
the system's proof theoretic complexity, but at the expense of
introducing the void logical rule of \textit{repetition} to balance
derivation trees.
\[
 \AxiomC{$\Gamma$}\RightLabel{$(\rep)$}\UnaryInfC{$\Gamma$}\DisplayProof
\]
Note that this rule is both logically valid and preserves the
sub-formula property, which in particular means that it does not harm 
computational tasks related to derivations as long as it does not occur 
too often.

It is well-known that, using $(\rep)$, the cut-elimination operator becomes 
a primitive recursive function which is continuous w.r.t.\ the standard 
metric on infinitary trees: 
the normalisation procedure requires only as
much information of the input as it produces output, using $(\rep)$ as
the last inference rule of the normal derivation, if the result cannot
immediately be determined (``please wait'').

In fact, associating some of the repetition rules with computation
steps bounds for the simply-typed lambda
calculus can be obtained that bound the sum of the number of
computation steps and the size of the
output~\cite{AehligJoachimski03}, strengthening earlier results by
Beckmann~\cite{Beckmann98}. 
Using Sch{\"u}tte's
$\omega$-rule~\cite{Schuette51} this method can also be applied to
G{\"o}del's~\cite{Goedel58} system $T$.

\bigskip

In this report, we will re-examine this situation.
We will show that the cut-reduction
operator can be understood as a polynomial time operation
natural way, see Observation~\ref{observation:feasibleCutReduction}.
We will work with proof notations which give implicit descriptions
of (infinite) propositional proofs:
a proof notation system will be a set which is equipped with some functions,
most importantly two which compute the following tasks:
\begin{itemize}
\item Given a notation $h$, compute the last inference $\tp(h)$ in the
denoted proof.
\item Given a notation $h$ and a number $i\in\NN$, compute a notation $h[i]$
for the $i$-th immediate sub-derivation of the derivation denoted by $h$.
\end{itemize}
Implicit proof notations given in this way uniquely determine a propositional
derivation tree, by exploring the derivation tree from its root and 
determining the inference at each node of the tree.
The cut-reduction operator will be defined on such implicitly described 
derivation trees. 
For this, we build on Buchholz' technical very smooth 
approach to notation systems for continuous 
cut-elimination~\cite{Buchholz91,buchholz:97}.
Our main result of the first part of the report
in particular implies the following statement, as can be seen from
Corollary~\ref{cor:2-k-1-bound-for-E-k-d}.
Let $2_n(x)$ denote the $n$-fold iteration of exponentiation $2^x$.
\begin{quote}
Let $d$ be some propositional derivation, and assume that all sub-proofs 
of $d$ can be denoted with notations of size bounded by $s$, and that
the height of $d$ is $h$.
Then, all sub-proofs of the derivation obtained from $d$ by reducing the 
complexity of cut-formulae by $k$ can be denoted by notations of size bounded 
by $2_{k-1}(2 h)\cdot s$.
\end{quote}
Observe that the size of notations is exponential only in the height of the
original derivation.
In the second part of this report  we will identify situations occurring in
proof-theoretical investigations of Bounded Arithmetic
where this height is bounded by an iterated
logarithm of some global size parameter, making these sizes feasible.

\bigskip

Bounded Arithmetic has been introduced by Buss \cite{buss:86} as theories of arithmetic
with a strong connection to computational complexity.
For sake of simplicity of this introduction, we will concentrate only
on the Bounded Arithmetic theories $\rS^i_2$ by Buss \cite{buss:86}.
These theories are given as first order theories of
arithmetic in a language which suitably extends that of Peano Arithmetic
where induction is restricted in two ways.
First, logarithmic induction is considered which only inducts over a logarithmic part 
of the universe of discourse.
\[
    \vhi(0)\land(\forall x)(\vhi(x)\imp\vhi(x+1))\imp(\forall x)\vhi(|x|) 
 \enspace. \]
Here, $|x|$ denotes the length of the binary representation of the natural number~$x$,
which defines a kind of logarithm on natural numbers.
Second, the properties which can be inducted on, must be described by a suitably
restricted (``bounded'') formula.
The class of formulae used here are the $\sib i$-formulae which exactly characterise
$\Si^p_i$, that is, properties of the $i$-th level of the polynomial time hierarchy of 
predicates.
The theory's $\rS^i_2$ main ingredients are the instances of logarithmic induction
for $\sib i$ formulae.

Let a (multi-)function $f$ be called $\sib j$-definable in $\rS^i_2$, if its
graph can be expressed by a $\sib j$-formula $\vhi$, such that the totality of
$f$, which renders as $(\forall x)(\exists y)\vhi(x,y)$, is provable from the
$\rS^i_2$-axioms in first-order logic.
The main results characterising definable (multi-) functions in Bounded Arithmetic
are the following.
\begin{itemize}
\item
Buss \cite{buss:86} has characterised the $\sib i$-definable functions of $\rS^i_2$ 
as $\FP^{\Si^b_{i-1}}$, the $i$-th level of the polynomial time hierarchy of 
functions.
\item
\krajicek \cite{krajicek:93} has characterised the 
$\sib{i+1}$-definable multi-functions of $S^i_2$ as the class
$\FP^{\Si^b_{i}}[wit,\OO{\log n}]$ of multi-functions which can
be computed in polynomial time using a witness oracle from $\Si^p_i$,
where the number of oracle queries is restricted to $\OO{\log n}$ many
($n$ being the length of the input).
\item
Buss and \krajicek \cite{buss:krajicek:94} have characterised the 
$\sib{i-1}$-definable multi-functions of $\rS^i_2$ as projections of solutions 
to problems from $\text{PLS}^{\Si^b_{i-2}}$, which is the class of 
polynomial local search problems relativised to $\Si^p_{i-2}$-oracles.
\end{itemize}

We will re-obtain all these definability characterisations by one unifying
method using the results from the first part of this report
in the following way.
First, we will define a suitable notation system \DerNBA for propositional 
derivations which are 
obtained by translating Bounded Arithmetic proofs.
The propositional translation used here is well-known in proof-theoretic investigations;
the translation has been described by Tait~\cite{Tait68},
and later was independently discovered by Paris and Wilkie~\cite{ParisWilkie85}.
In the Bounded-Arithmetic world it is known as the \emph{Paris-Wilkie translation.}

Applying the machinery from the first part we obtain a notation system
\ceDerNBA of cut-elimination for \DerNBA.
\ceDerNBA will have the property that its implicit descriptions, most notably
the functions $\tp(h)$ and $h[i]$ mentioned above, will be polynomial time
computable.

This allows us to formulate a general local search problem on \ceDerNBA which
is suitable to characterise definable multi-functions for Bounded Arithmetic.
Assume that $(\forall x)(\exists y)\vhi(x,y)$, describing the totality of some 
multi-function, is provable in some Bounded Arithmetic theory.
Fix a particularly nice formal proof $p$ of this.
Given $N\in\NN$ we want to describe a procedure which 
finds some $K$ such that $\vhi(\ul N,\ul K)$ holds.
Invert the proof $p$ of
$(\forall x)(\exists y)\vhi(x,y)$ to a proof of 
$(\exists y)\vhi(x,y)$ where $x$ is fresh a variable,
then substitute $\ul N$ for all occurrences of $x$. This yields a
proof of $(\exists y)\vhi(\ul N, y)$. Adding an appropriate number of
cut-reduction operators we obtain a proof with all cut-formulae of (at
most) the same logical complexity as $\vhi$. It should be noted that a
notation $h(N)$ for this proof can be computed in time polynomial in
$N$.

The general local search problem which finds a witness for 
$(\exists y)\vhi(\ul N,y)$ can now be characterised as follows.
Its instance is given by $N$.
The set of solutions are those notations of a suitable size,
which denote a derivation having the property that the derived sequent
is equivalent to $(\exists y)\vhi(\ul N,y)\lor\psi_1\lor\dots\lor\psi_l$
where all $\psi_i$ are ``simple enough''
and false.
An initial solution is given by $h(N)$.
A neighbour to a solution $h$ is a solution
which denotes an immediate sub-derivation of the derivation denoted by $h$,
if this exists, and $h$ otherwise.
The cost of a notation is the height of the denoted derivation.
The search task is to find a notation in the set of solutions which is a 
fixpoint of the neighbourhood function.
Obviously, a solution to the search task must exist. In fact, any
solution of minimal cost has this property.
Now consider any solution to the search problem. It must have the
property, that none of the immediate sub-derivations is in the solution
space. This can only happen if the last inference derives 
$(\exists y)\vhi(\ul N,y)$ from a true statement $\vhi(\ul N,\ul K)$
for some $K\in\NN$.
Thus $K$ is a witness to $(\exists y)\vhi(\ul N,y)$, and we can output $K$ 
as a solution to our original witnessing problem.

Depending on the complexity of logarithmic induction present in the  
Bounded Arithmetic theory we started with, and the level of definability,
we obtain local search problems defined by functions of some level of the 
polynomial time hierarchy, and different bounds to the cost function.
For example, if we start with the $\sib i$-definable functions of $\rS^i_2$,
we obtain a local search problem defined by properties in $\FP^{\sib{i-1}}$, 
where the cost function is bounded by $|N|^{\OO{1}}$.
Thus, by following the canonical path through the search problem which
starts at the initial value and iterates the neighbourhood function, we
obtain a path of polynomial length, which describes a procedure
in $\FP^{\sib{i-1}}$ to compute a witness.

\bigskip

Other research related to our investigations is a paper by 
Buss \cite{buss:04} which also makes use of the Paris-Wilkie translation to
obtain witnessing results by giving uniform descriptions of translated 
proofs.
However, Buss' approach does not explicitely involve cut-elimination.
Dynamic ordinal analysis \cite{beckmann:doa:03,beckmann:gdoa:06} characterises
the heights of propositional proof trees obtained via the Paris-Wilkie 
translation and cut-reduction.
Therefore, it is not surprising that the bounds obtained by 
dynamic ordinal analysis coincide with the bounds on cost functions we are 
exploiting here.

The potential of our approach to the characterisations of definable search 
problems via notation systems is that it may lead to characterisations
of so far uncharacterised definable search problems, most notably the
$\sib1$-definable search problems in $\rS^i_2$ for $i\ge 3$.
\end{section}

\begin{section}{Proof Systems}

Let $S$ be a set.
The set of all subsets of $S$ will be denoted by $\pow(S)$,
the set of all finite subsets of $S$ will be denoted by $\finpow(S)$.

\begin{DEF}[sequent]
Let $\cF$ be a set (of \emph{formulae}), 
$\ieq$ a binary relation on $\cF$ (\emph{identity between formulae}),
and $\rk\colon\pow(\cF)\times\cF\to\NN$ a function (\emph{rank}).
A \emph{sequent} over $\cF,\ieq,\rk$ is a finite subset of $\cF$.
We use $\Ga,\De,\dots$ as syntactic variables to denote sequents.
With $\seqieq{\De}$ we denote the set 
$\setof{A\in\cF}{(\exists B\in\De)A\ieq B}$.
\end{DEF}

We usually write $A_1,\dots,A_n$ for
$\{A_1,\dots,A_n\}$ and $A,\Ga,\De$ for $\{A\}\cup\Ga\cup\De$, etc.
We always write $\Crk(A)$ instead of $\rk(\cC,A)$.

We repeat standard Buchholz notation for proof systems~\cite{buchholz:97}. 

\begin{DEF}
A \emph{proof system $\PrSys$ over $\cF,\ieq,\rk$}\: is given by
\begin{itemize}
\item a set of formal expressions called \emph{inference symbols} 
(syntactic variable $\IS$);
\item for each inference symbol $\IS$ an ordinal $|\IS|\le\om$, a sequent 
$\De(\IS)$ and a family of sequents $(\De_\iota(\IS))_{\iota<|\IS|}$.
\end{itemize}
\end{DEF}

Proof systems may have inference symbols of the form $\Cut_C$ for
$C\in\cF$; these are called ``cut inference symbols'' and their use
will (in Definition~\ref{def:quasi-derivations}) be measured by the $\cC$-cut rank.

\begin{notation}
By writing 
\AxiomC{$\dots\De_\iota\dots(\iota< I)$}
\LeftLabel{$(\IS)$}
\UnaryInfC{$\De$}
\DisplayProof
we declare $\IS$ as an inference symbol with $|\IS|=I$, $\De(\IS)=\De$,
$\De_\iota(\IS) = \De_\iota$.
If $|\IS|=n$ we write
\AxiomC{$\De_0\ \De_1\ \dots\ \De_{n-1}$}
\UnaryInfC{$\De$}
\DisplayProof
instead of
\AxiomC{$\dots\De_\iota\dots(\iota< I)$}
\UnaryInfC{$\De$}
\DisplayProof.
\end{notation}

\begin{DEF}[Inductive definition of $\PrSys$-quasi derivations]%
\label{def:quasi-derivations}
If $\IS$ is an inference symbol of $\PrSys$, 
and $(d_\iota)_{\iota<|\IS|}$ is a sequence of $\PrSys$-quasi derivations,
then $d:=\IS(d_\iota)_{\iota<|\IS|}$
is an \emph{$\PrSys$-quasi derivation} with
\begin{align*}
\Ga(d) &:= \De(\IS)\cup
   \bigcup_{\iota<|\IS|}(\Ga(d_\iota)\setminus\seqieq{\De_\iota(\IS)})
   && \text{(\emph{endsequent of $d$})} \\
\last(d) &:= \IS
   && \text{(\emph{last inference of $d$})} \\
d(\iota) &:= d_\iota \text{ for }\iota<|\IS|
   && \text{(\emph{sub-derivation})} \\
\Ccrk(d) &:=  \sup(\{\Crk(\IS)\}\cup\setof{\Ccrk(d_\iota)}{\iota<|\IS|})
   && \text{(\emph{cut-rank of $d$})} \\
     & \text{where } \Crk(\IS):=
 \begin{cases}
   \Crk(C)+1 & \text{if } \IS=\Cut_C \\
   0 & \text{otherwise}
 \end{cases}  \\
\hgt(d) &:= \sup\setof{\hgt(d_\iota)+1}{\iota<|\IS|} 
   && \text{(\emph{height of $d$})} \\
\dsz{d} &:= (\sum_{\iota<|\IS|}\dsz{d_\iota)}+1
   && \text{(\emph{size of $d$})}
\end{align*}
\end{DEF}

\end{section}

\begin{section}{The infinitary proof system}

\begin{DEF}
Let $\infConn=\{\verum,\falsum,\tbw,\tbv\}$ be the set of (symbols for)
connectives for infinitary logic. 
Their arity is given by $|\verum|=|\falsum|=0$ and
$|\tbw|=|\tbv|=\om$.
We define a negation of the connectives according to the de Morgan laws:
$\neg(\verum)=\falsum$, $\neg(\falsum)=\verum$,
$\neg(\tbw)=\tbv$, and $\neg(\tbv)=\tbw$.
\end{DEF}

\newcommand{\myop}{c}
\begin{DEF}
The set of all infinitary formulae $\cL_\infty$ together with their rank 
is inductively defined by the clause:
if $\myop\in\infConn$ and $A_\iota\in\cL_\infty$ for $\iota<|\myop|$ 
then $\myop(A_\iota)_{\iota<|\myop|}\in\cL_\infty$
and $\Crk(\myop(A_\iota)_{\iota<|\myop|})=\sup_{\iota<|\myop|}(\Crk(A_\iota)+1)$.
\end{DEF}

\noi
\textbf{Notation}\\
We denote $\verum()$ by $\verum$ and $\falsum()$ by $\falsum$.

\smallskip

\begin{DEF}
$\neg$ denotes the operation on $\cL_\infty$ which computes 
negation according to the de Morgan rules, i.e.
\[
  \neg\left(\myop(A_\iota)_{\iota<|\myop|}\right) :=
   \neg(\myop)\big(\neg(A_\iota)\big)_{\iota<|\myop|}
\]
\end{DEF}

\begin{DEF}
The set of all infinitary formulae of finite rank is denoted with \Finf.
The identity between \Finf-formulae is the ``true'' set-theoretic equality.
\end{DEF}

\begin{DEF}\label{def:infinitary_proof_system}
The \emph{infinitary proof system $\PrSysInf$} is 
the proof system over $\Finf$ which is given by the 
following set of inference symbols:\\[3ex]
\AXC{\mbox{}}
\LL{$(\Ax)$\quad}
\UIC{$\verum$}
\DP
\\[2ex]
\AXC{\dots\quad$A_\iota$\quad\dots\quad$(\iota<\om)$}
\LL{$(\tbw_A)$\quad}
\UIC{$A$}
\DP
for $A=\tbw(A_\iota)_{\iota<\om}\in\Finf$
\\[2ex]
\AXC{$A_i$}
\LL{$(\tbv^{i}_A)$\quad}
\UIC{$A$}
\DP
for $A=\tbv(A_\iota)_{\iota<\om}\in\Finf$ and $i<\om$
\\[2ex]
\AXC{$C$}
\AXC{$\neg C$}
\LL{$(\Cut_{C})$\quad}
\BIC{$\emptyset$}
\DP
for $C\in\Finf$
\\[2ex]
\AXC{$\emptyset$}
\LL{$(\Rep)$\quad}
\UIC{$\emptyset$}
\DP
\end{DEF}

\begin{DEF}
The \emph{$\PrSysInf$-derivations} are the $\PrSysInf$-quasi derivations.
\end{DEF}

With a $\PrSysInf$-derivation $d=\IS(d_\iota)_{\iota<|\IS|}$ we can
associate a function from $\NN^{<\om}$ to $\PrSysInf$ by letting
$d(\seq{}):=\last(d)$ and
\[
 d(\seq{i}\frown s) := \begin{cases}
  d_i(s) & \text{if } i<|\IS| \\
  \Ax & \text{otherwise}
 \end{cases}
\]

\end{section}

\begin{section}{Notation system for infinitary formulae}

\begin{DEF}\label{def:NotationSystemForFormulas}
A \emph{notation system for (infinitary) formulae} is a set $\ForN$ of
``formulae'', together with 
four functions
$\tp\colon\ForN\to\{\verum,\falsum,\tbw,\tbv\}$,
$\cdot[\cdot]\colon\ForN\times\NN\to\ForN$, 
$\neg\colon\ForN\to\ForN$, and
$\rk\colon\pow(\ForN)\times\ForN\to\NN$
called ``outermost connective'', ``sub-formula'', ``negation'' and ``rank'', 
and a relation 
$\ieq\,\,\subseteq\ForN\times\ForN$
called ``intensional equality'', 
such that
$\tp(\neg(f))=\neg(\tp(f))$, 
$\neg(f)[n]=\neg(f[n])$, 
$\Crk(f)=\Crk(\neg f)$,
$\Crk(f[n])<\Crk(f)$ for $n<|\tp(f)|$, and
$f\ieq g$ implies $\tp(f)=\tp(g)$, $f[n]\ieq g[n]$, 
$\neg(f)\ieq\neg(g)$ and $\Crk(f)=\Crk(g)$.
\end{DEF}

It should be noted that if $\ForN$ is a notation system for formulae,
then so is $\ForN/\ieq$ in the obvious way; moreover, in $\ForN/\ieq$
the intensional equality is true equality in the quotient. The reason
why we nevertheless explicitly consider an (intensional) equality
relation is that we are
interested in the computational complexity of notation systems and
therefore prefer to take notations as the strings that arise
naturally, rather than working on the quotient. Note that the latter would
require us to compute canonical representations anyway and so would
just push the problem to a different place.

It should also be noted that the intensional equality is truly intensional. Two
formulae are only equal, if they are given to us as being equal. 
The obvious extensional equality would be the largest bisimulation, 
that is, the largest relation $\sim\subset\ForN\times\ForN$ satisfying
$f\sim g\to \tp(f)=\tp(g)\land f[n]\sim g[n]\land \Crk(f)=\Crk(g)
\land \neg f\sim\neg g$. 
However, as most extensional concepts, the largest
bisimulation is undecidable in almost all interesting cases and
therefore not suited for an investigation of effective notations. 

\begin{DEF}
Let $\ForN=(\ForN,\tp,\cdot[\cdot],\rk,\ieq)$ be a notation system for 
infinitary formulae.
The \emph{interpretation $\ieval{f}$ of $f\in\ForN$} is inductively defined as
\[ \ieval{f} = \tp(f)(\ieval{f[\iota]})_{\iota<|\tp(f)|} \]
\end{DEF}

\begin{observation}\label{obs:ieval-fla}
The following properties hold.
\begin{enumerate}
\item
  $f\sim g \Gdw \ieval{f}=\ieval{g}$,
\item
  $f\ieq g \Imp \ieval{f}=\ieval{g}$.
\end{enumerate}
\end{observation}

\end{section}

\begin{section}{Semiformal proof systems}

Let $\ForN=(\ForN,\tp,\cdot[\cdot],\rk,\ieq)$ be a notation system for 
infinitary formulae.

\begin{DEF}\label{def:semiformal_proof_systems}
The \emph{semiformal proof system $\PrSysForN$ over $\ForN$} 
is the proof system over $\ForN$
which is given by the following set of inference symbols:\\[3ex]
\AXC{\mbox{}}
\LL{$(\Ax_A)$\quad}
\UIC{$A$}
\DP
for $A\in\ForN$ with $\tp(A)=\verum$
\\[2ex]
\AXC{\dots\quad$C[n]$\quad\dots\quad$(n\in\NN)$}
\LL{$(\tbw_C)$\quad}
\UIC{$C$}
\DP
for $C\in\ForN$ with $\tp(C)=\tbw$
\\[2ex]
\AXC{$C[i]$}
\LL{$(\tbv^{i}_C)$\quad}
\UIC{$C$}
\DP
for $C\in\ForN$ with $\tp(C)=\tbv$ and $i\in\NN$
\\[2ex]
\AXC{$C$}
\AXC{$\neg C$}
\LL{$(\Cut_{C})$\quad}
\BIC{$\emptyset$}
\DP
for $C\in\ForN$ with $\tp(C)\in\{\verum,\tbw\}$
\\[2ex]
\AXC{$\emptyset$}
\LL{$(\Rep)$\quad}
\UIC{$\emptyset$}
\DP
\end{DEF}

\bigskip

\noi
\textbf{Abbreviations}\\
For $\tp(C)\in\{\falsum,\tbv\}$ let
\AXC{$C$}
\AXC{$\neg C$}
\LL{$(\Cut_{C})$\ }
\BIC{$\emptyset$}
\DP
denote
\AXC{$\neg C$}
\AXC{$C$}
\LL{$(\Cut_{\neg C})$\ }
\BIC{$\emptyset$}
\DP.

\begin{DEF}
The \emph{$\PrSysForN$-derivations} are the $\PrSysForN$-quasi derivations.
\end{DEF}

Later in our applications, we will be concerned only with derivations
of finite height, for which we can formulate slightly sharper upper
bounds on cut-reduction than in the general (infinite) case 
($2^\al$ versus $3^\al$).
Thus, from now on we will restrict attention to derivations of 
finite height only.

\begin{DEF}\label{def:turnstile}
Let $d\vdash^\al_{\cC,m}\Ga$ denote that $d$ is an $\PrSysForN$-derivation with
$\Ga(d)\subseteq\seqieq\Ga$, $\Ccrk(d)\le m$, and $\hgt(d)\le\al<\om$\enspace.
\end{DEF}

\begin{DEF}
The \emph{interpretation $\ieval{d}$ of a 
$\PrSysForN$-derivation $d=\IS(d_\iota)_{\iota<|\IS|}$} is defined as
\[  \ieval{d} := \ieval{\IS} (\ieval{d_\iota})_{\iota<|\IS|} \]
where $\ieval{\IS}$ is defined by
\begin{align*}
\ieval{\Ax_A} &:= \Ax \\
\ieval{\tbw_A} &:= \tbw_\ieval{A} \\
\ieval{\tbv^i_A} &:= \tbv^i_\ieval{A} \\
\ieval{\Cut_C} &:= \Cut_\ieval{C} \\
\ieval{\Rep} &:= \Rep
\end{align*}
\end{DEF}

\begin{observation}
$\Ga(\ieval d)\subseteq\ieval{\Ga(d)}$
\end{observation}
\begin{proof}
Induction on $d$. The ``$\subseteq$'', instead of the expected ``$=$''
is due to the fact, that only formulae are removed from the conclusion
that are intensionally equal; compare also
Observation~\ref{obs:ieval-fla}.
\end{proof}

\end{section}

\begin{section}{Cut elimination for semiformal systems}

Let $\ForN=(\ForN,\tp,\cdot[\cdot],\rk,\ieq)$ be a notation system for 
infinitary formulae, 
and $\PrSysForN$ the semiformal proof system over $\ForN$.
We define Mints' continuous cut-reduction operator \cite{Mints75,KreiselMintsSimpson75}
following the description given by Buchholz \cite{Buchholz91}.
The only modification is our explicit use of intensional equality.

\begin{Th}[and Definition]\label{th:I-operator}
Let $C\in\ForN$ with $\tp(C)=\tbw$, and $k<\om$ be given.
We define an operator $\opI^k_C$ such that:\qquad
$d\vdash^\al_{\cC,m}\Ga,C$ 
\Imp $\opI^k_C(d)\vdash^{\al}_{\cC,m}\Ga,C[k]$.
\end{Th}

\begin{proof}[Proof by induction on the build-up of $d$:]
W.l.o.g.\ we may assume that $\Ga=\Ga(\derd)\setminus\seqieq{\{C\}}$.

\medskip

\noi
\textbf{Case 1.}
$\last(\derd)\in\setof{\tbw_D}{D\ieq C}$.
Then
\[
  \opI^k_C(\derd) := \Rep (\opI^k_C(\derd(k))) 
\]
is a derivation as required.

\medskip

\noi
\textbf{Case 2.}
$\IS:=\last(\derd)\notin\setof{\tbw_D}{D\ieq C}$.
Then
\[
  \opI^k_C(\derd) := \IS (\opI^k_C(\derd(i)))_{i<|\IS|}
\]
is a derivation as required.
\end{proof}

\begin{Th}[and Definition]\label{th:R-operator}
Let $C\in\ForN$ with $\tp(C)\in\{\verum,\tbw\}$ be given.
We define an operator $\opR_{C}$ such that:\qquad
$\derd_0\vdash^\al_{\cC,m}\Ga,C$ \Und
$\derd_1\vdash^\be_{\cC,m}\Ga,\neg C$ \Und
$\Crk(C)\le m$ 
\Imp $\opR_{C}(\derd_0,\derd_1)\vdash^{\al+\be}_{\cC,m}\Ga$.
\end{Th}

\begin{proof}[Proof by induction on the build-up $d$:]
W.l.o.g.\ we may assume that 
$\Ga=(\Ga(d_0)\setminus\seqieq\{C\})\cup(\Ga(d_1)\setminus\seqieq\{\neg C\})$.
Let $\IS=\last(d_1)$.

\medskip

\noi
\textbf{Case 1.}
$\De(\IS)\cap\seqieq{\{\neg C\}}=\emptyset$.
Then $\De(\IS)\subseteq \Ga$ and $d_1(i)\vdash^{\be_i}_{\cC,m} \Ga,\neg C,\De_i(\IS)$
with $\be_i<\be$ for all $i<|\IS|$.
By induction hypothesis we obtain
$\opR_{C}(\derd_0,\derd_1(i))\vdash^{\al+\be_i}_{\cC,m} \Ga,\De_i(\IS)$ 
for $i<|\IS|$.
Hence
\[
  \opR_{C}(\derd_0,\derd_1) := 
    \IS (\opR_{C}(\derd_0,\derd_1(i)))_{i<|\IS|}
\]
is a derivation as required.

\medskip

\noi
\textbf{Case 2.}
$\De(\IS)\cap\seqieq{\{\neg C\}}\neq\emptyset$.
Then $\tp(C)\neq\verum$, because otherwise 
there is some $D\in\De(\IS)$ with $\tp(D)=\falsum$,
but this is not satisfied by any of the inference symbols 
of the semiformal system $\PrSysForN$.
Hence $\tp(C)=\tbw$.
We obtain that $\IS=\tbv^k_{D}$ for some $k\in\NN$ and $D\ieq \neg C$,
and $d_1(0)\vdash^{\be_0}_{\cC,m} \Ga,\neg C,\neg C[k]$ with $\be_0<\be$.
By induction hypothesis we obtain
$\opR_{C}(\derd_0,\derd_1(0))\vdash^{\al+\be_0}_{\cC,m} \Ga,\neg C[k]$.
The Inversion Theorem shows
$\opI^k_C(d_0)\vdash^\al_{\cC,m} \Ga, C[k]$.
Now $\Crk(C[k])<\Crk(C)\le m$, hence
\[
  \opR_{C}(\derd_0,\derd_1) := 
    \Cut_{C[k]}(\opI^k_C(d_0),\opR_{C}(\derd_0,\derd_1(0)))
\]
is a derivation as required.
\end{proof}

\begin{Th}[and Definition]\label{th:E-operator}
We define an operator $\opE$ such that:\\
$\derd\vdash^\al_{\cC,m+1}\Ga$ \Imp 
$\opE(\derd)\vdash^{2^\al-1}_{\cC,m}\Ga$.
\end{Th}

\begin{proof}[Proof by induction on the build-up of $d$:]
W.l.o.g.\ we may assume that $\Ga=\Ga(\derd)$.

\medskip

\noi
\textbf{Case 1.}
$\last(\derd)=\Cut_{C}$.
Then $\Crk(C)\le m$ and 
$\derd(0)\vdash^{\al_0}_{\cC,m+1}\Ga,C$
and
$\derd(1)\vdash^{\al_0}_{\cC,m+1}\Ga,\neg C$
with $\al_0<\al$.
By induction hypothesis we obtain
$\opE(\derd(0))\vdash^{2^{\al_0}-1}_{\cC,m}\Ga,C$
and
$\opE(\derd(1))\vdash^{2^{\al_0}-1}_{\cC,m}\Ga,\neg C$.

\medskip

\noi
\textbf{Case 1.1.}
$\tp(C)\in\{\verum,\tbw\}$, then by the last Theorem
$\opR_{C}(\opE(\derd(0)),\opE(\derd(1)))
 \vdash^{2\cdot 2^{\al_0}-2}_{\cC,m}\Ga$, and
\[
  \opE(\derd) := \Rep (\opR_{C}(\opE(\derd(0)),\opE(\derd(1))))
\]
is a derivation as required.

\medskip

\noi
\textbf{Case 1.2.}
$\tp(C)\notin\{\verum,\tbw\}$, then 
$\opR_{\neg C}(\opE(\derd(1)),\opE(\derd(0)))
  \vdash^{2\cdot 2^{\al_0}-1}_{\cC,m}\Ga$.
Continue as before.

\medskip

\noi
\textbf{Case 2.}
$\IS:=\last(\derd)\neq\Cut_C$.  
Then
\[
  \opE(\derd) := \IS (\opE(\derd(i)))_{i<|\IS|}
\]
is as required.
\end{proof}

\begin{REM}
Immediately from the definition we note that the operators $\opI$,
$\opR$, and $\opE$ only inspects the last inference symbol of a
derivation to obtain the last inference symbol of the transformed
derivation. It should be noted that this continuity would not be
possible without the repetition rule.
\end{REM}

\end{section}

\begin{section}{Notations for derivations and cut-elimination}

Let $\ForN$ be a notation system for formulae,
and $\PrSysForN$ the semiformal proof system over $\ForN$
from Definition~\ref{def:semiformal_proof_systems}.

\begin{DEF}\label{def:notation-for-proof-system}
A \emph{notation system for $\PrSysForN$} is 
a set $\DerN$ of \emph{notations} and functions
$\tp\colon\DerN\to\PrSysForN$,
$\cdot[\cdot]\colon\DerN\times\NN\to\DerN$, 
$\Ga\colon\DerN\to\finpot(\ForN)$,
$\crk\colon\pow(\ForN)\times\DerN\to\NN$,
and
$\ord,\sz{\cdot}\colon\DerN\to\NN\setminus\{0\}$
called \emph{denoted last inference},
\emph{denoted sub-derivation},
\emph{denoted end-sequent},
\emph{denoted cut-rank},
\emph{denoted height} and \emph{size},
such that
$\Ccrk(h[n])\le\Ccrk(h)$,
$\tp(h)=\Cut_C$ implies $\Crk(C)<\Ccrk(h)$,
$\ord(h[n])<\ord(h)$ for $n<|\tp(h)|$,
and the following local faithfulness property holds for $h\in\DerN$:
\[  
\De(\tp(h))\cup \bigcup_{\iota<|\tp(h)|} 
        \Big(\Ga(h[\iota])\setminus\seqieq{\De_\iota(\tp(h))}) \Big)
\subseteq \seqieq{\Ga(h)}
\enspace. \]
\end{DEF}

\begin{prop}\label{prop:cons-faithfull}
\[  \Ga(h[j])\subseteq \seqieq{\Big(\Ga(h)\cup\De_j(\tp(h))\Big)}  \]
\end{prop}

\begin{DEF}
Let $\DerN=(\DerN,\tp,\cdot[\cdot],\ord,\sz{\cdot})$ be a 
notation system for $\PrSysForN$ .
The \emph{interpretation $\eval{h}$ of $h\in\DerN$} is inductively defined 
as the following $\PrSysForN$-derivation:
\[  \eval{h} := \tp(h) (\eval{h[n]})_{n<|\tp(h)|} \]
\end{DEF}

\begin{observation}\label{observ:formal-semiformal}
For $h\in\DerN$ we have
\begin{align*}
  \last(\eval{h}) &= \tp(h)  \\
  \eval{h}(\iota) &= \eval{h[\iota]}  \quad\text{ for } \iota<|\tp(h)| \\
  \Ga(\eval{h}) &\subseteq \seqieq{\Ga(h)}
\end{align*}
\end{observation}

We now extend a notation system $\DerN$ for $\PrSysForN$ to notation system 
for cut-elimination on $\DerN$, by adding notations for the operators
$\opI$, $\opR$ and $\opE$ from the previous section.

\begin{DEF}
The \emph{notation system $\ceDerN$ for cut-elimination on $\DerN$} is given by 
the set of terms $\ceDerN$ which are inductively defined by
\begin{itemize}
\item $\DerN\subset\ceDerN$,
\item
 $h\in\ceDerN$, $C\in\ForN$ with $\tp(C)=\tbw$, $k<\om$
   \Imp $\I^k_C h\in\ceDerN$,
\item
 $h_0,h_1\in\ceDerN$, $C\in\ForN$ with $\tp(C)\in\{\verum,\tbw\}$
   \Imp $\R_{C} h_0 h_1\in\ceDerN$,
\item
 $h\in\ceDerN$ \Imp $\E h\in\ceDerN$,
\end{itemize}
where $\I,\R,\E$ are new symbols,
and functions
$\tp\colon\ceDerN\to\PrSysForN$,
$\cdot[\cdot]\colon\ceDerN\times\NN\to\ceDerN$,
$\Ga\colon\ceDerN\to\finpot(\ForN)$,
$\crk\colon\pow(\ForN)\times\ceDerN\to\NN$,
$\ord\colon\ceDerN\to\NN\setminus\{0\}$ and
$\sz{\cdot}\colon\ceDerN\to\NN$
defined by recursion on the build-up of $h\in\ceDerN$:
\begin{itemize}
\item 
If $h\in\DerN$ then all functions are inherited from $\DerN$.

\item 
$h=\I^k_C h_0$: 
Let $\Ga(h):=\{C[k]\}\cup(\Ga(h_0)\setminus\seqieq{\{C\}})$,
$\Ccrk(h):=\Ccrk(h_0)$,
$\ord(h):=\ord(h_0)$, and $|h|:=|h_0|+1$.

\smallskip

\noi
\textbf{Case 1.}
$\tp(h_0)\in\setof{\tbw_D}{D\ieq C}$.
Then let $\tp(h):=\Rep$, and $h[0]:=\I^k_C h_0[k]$.

\smallskip

\noi
\textbf{Case 2.}
Otherwise, let $\tp(h):=\tp(h_0)$, and $h[i]:=\I^k_C h_0[i]$.

\item
$h=\R_{C} h_0 h_1$:
Let $\IS:=\tp(h_1)$.
We define $\Ga(h):=(\Ga(h_0)\setminus\seqieq{\{C\}})\cup
 (\Ga(h_1)\setminus\seqieq{\{\neg C\}})$,
$\Ccrk(h):=\max\{\Ccrk(h_0),\Ccrk(h_1)\}$,
$\ord(h):=\ord(h_0)+\ord(h_1)$, and $|h|:=|h_0|+|h_1|+1$.
\smallskip

\noi
\textbf{Case 1.}
$\De(\IS)\cap\seqieq{\{\neg C\}}=\emptyset$:
Then let $\tp(h):=\IS$, and $h[i]:=\R_{C} h_0 h_1[i]$.

\smallskip

\noi
\textbf{Case 2.}
Otherwise, $\tp(C)\neq\verum$, because if not 
there would be some $D\in\De(\IS)$ with $\tp(D)=\falsum$,
but this is not satisfied by any of the inference symbols 
of the semiformal system $\PrSysForN$.
Hence $\tp(C)=\tbw$.
Thus $\IS=\tbv^k_{D}$ for some $k\in\NN$ and $D\ieq \neg C$.
Then let $\tp(h):=\Cut_{C[k]}$ and 
$h[0]:=\I^k_C h_0$, $h[1]:=\R_{C} h_0 h_1[0]$.

\item
$h=\E h_0$:
Let $\Ga(h):=\Ga(h_0)$,
$\Ccrk(h):=\Ccrk(h_0)\aminus1$,
$\ord(h):= 2^{\ord(h_0)}-1$, and $|h|:=|h_0|+1$.

\smallskip

\noi
\textbf{Case 1.}
$\tp(h_0)=\Cut_{C}$:
Then let $\tp(h):=\Rep$ and \\
let $h[0]:=\R_{C}\E h_0[0]\E h_0[1]$ if $\tp(C)\in\{\verum,\tbw\}$,\\
let $h[0]:=\R_{\neg C}\E h_0[1]\E h_0[0]$ if $\tp(C)\notin\{\verum,\tbw\}$.

\smallskip

\noi
\textbf{Case 2.}
Otherwise, let $\tp(h):=\tp(h_0)$, and $h[i]:=\E h_0[i]$.

\end{itemize}
\end{DEF}

\begin{proof}
The just defined system is a notation system for $\PrSysForN$
in the sense of Definition~\ref{def:notation-for-proof-system}.
To prove this we have to show that 
\begin{equation}\label{eqCutElimNotationProof}
  \ord(h[n])<\ord(h)  \qquad\text{for}\qquad n<|\tp(h)|
\end{equation}
and that the local faithfulness property for $\Ga$ holds.
We start by proving \eqref{eqCutElimNotationProof}
by induction on the build-up of $h\in\ceDerN$.

If $h\in\DerN$ then \eqref{eqCutElimNotationProof} is inherited from $\DerN$.
If $h=\I^k_C h_0$ then $h[n]=\I^k_C h_0[n']$ for some $n'$ and 
\eqref{eqCutElimNotationProof} is immediate by induction hypothesis.

Now let us consider the case $h=\R_Ch_0h_1$.
If $h[n]=\R_Ch_0h_1[n']$ for some $n'$ then \eqref{eqCutElimNotationProof} 
is immediate by induction hypothesis.
The other case is that $h[0]=\I^k_Ch_0$ for some $k$.
We compute
\[ \ord(h[0]) = \ord(\I^k_Ch_0) = \ord(h_0) < \ord(h_0)+\ord(h_1) = \ord(h) \]
since $\ord(h_1)>0$.

Finally, let us consider the case $h=\E h_0$.
If $h[n]=\E h_0[n]$ then \eqref{eqCutElimNotationProof} is immediate by 
induction hypothesis.
Otherwise, we are in the case $h[0]=\R_C(\E h_0[i])(\E h_0[j])$ for
some $C,i,j$.
By induction hypothesis we obtain that $\ord(h_0[i])\le\ord(h_0)-1$
and $\ord(h_0[j])\le\ord(h_0)-1$.
Hence
\begin{align*}
 \ord(\R_C(\E h_0[i])(\E h_0[j]))
  &= \ord(\E h_0[i])+\ord(\E h_0[j])
  = 2^{\ord(h_0[i])}-1+2^{\ord(h_0[j])}-1 \\
  &< 2\cdot2^{\ord(h_0)-1}-1
  = 2^{\ord(h_0)}-1 = \ord(h)
\end{align*}

We now turn to the local faithfulness property of $\Ga$ which we also 
prove by induction on the build-up of $h\in\ceDerN$.
We abbreviate
\[  
*(h) \quad:=\quad
   \De(\tp(h))\cup \bigcup_{\iota<|\tp(h)|} 
        \Big(\Ga(h[\iota])\setminus\seqieq{\De_\iota(\tp(h))}) \Big)
\enspace, \]
then we have to show $*(h)\subseteq\seqieq{\Ga(h)}$.

\begin{itemize}
\item 
If $h\in\DerN$ then the local faithfulness property is inherited from $\DerN$.

\item 
If $h=\I^k_C h_0$, 
then $\Ga(h):=\{C[k]\}\cup(\Ga(h_0)\setminus\seqieq{\{C\}})$.

\smallskip

\noi
\textbf{Case 1.}
$\tp(h_0)\in\setof{\tbw_D}{D\ieq C}$.
Then $\Ga(h_0[k]) \subseteq *(h_0)\cup\seqieq{\{C[k]\}}$
hence
\begin{align*}
*(h) &= \emptyset\cup\Ga(\I^k_C h_0[k]) \\
 &= \{C[k]\}\cup\Big( \Ga(h_0[k]) \setminus\seqieq{\{C\}}\Big) \\
 &\subseteq
    \{C[k]\}\cup\Big( *(h_0) \setminus\seqieq{\{C\}}\Big) \\
 &\stackrel{i.h.}{\subseteq}
    \{C[k]\}\cup\Big( \seqieq{\Ga(h_0)} \setminus\seqieq{\{C\}}\Big) 
 \quad\subseteq\quad \seqieq{\Ga(h)}
\end{align*}

\smallskip

\noi
\textbf{Case 2.}
Otherwise, we compute
\begin{align*}
*(h) &= \De(\tp(h_0))\cup\bigcup_{\iota<|\tp(h_0)|}
     \Big(\Ga(\I^k_C h_0[\iota])\setminus\seqieq{\De_\iota(\tp(h_0))}\Big) \\
 &=
   \De(\tp(h_0))\cup \bigcup_{\iota<|\tp(h_0)|}\Big(
     \Big[\{C[k]\}\cup\big(\Ga(h_0[\iota])\setminus\seqieq{\{C\}}\big)\Big]
           \setminus\seqieq{\De_\iota(\tp(h_0))}  \Big) \\
 &\subseteq
   \{C[k]\}\cup\Big(
     \Big[\De(\tp(h_0))\cup \bigcup_{\iota<|\tp(h_0)|}
       \big( \Ga(h_0[\iota])\setminus\seqieq{\De_\iota(\tp(h_0))} \big) \Big]
     \setminus\seqieq{\{C\}}\Big) \\
 &=
   \{C[k]\}\cup\Big(
     *(h_0)
     \setminus\seqieq{\{C\}}\Big) \\
 &\stackrel{i.h.}{\subseteq}
   \{C[k]\}\cup\Big(
     \seqieq{\Ga(h_0)}
     \setminus\seqieq{\{C\}}\Big) 
 \quad\subseteq\quad \seqieq{\Ga(h)}
\end{align*}

\item
$h=\R_{C} h_0 h_1$:
Let $\IS:=\tp(h_1)$.
We have $\Ga(h):=(\Ga(h_0)\setminus\seqieq{\{C\}})\cup
 (\Ga(h_1)\setminus\seqieq{\{\neg C\}})$.
\smallskip

\noi
\textbf{Case 1.}
$\De(\IS)\cap\seqieq{\{\neg C\}}=\emptyset$:
We compute
\begin{align*}
*(h) &= \De(\IS)\cup\bigcup_{\iota<|\IS|}
     \Big(\Ga(\R_C h_0 h_1[\iota])\setminus\seqieq{\De_\iota(\IS)}\Big) \\
 &=
   \De(\IS)\cup \bigcup_{\iota<|\IS|}\Big(
     \Big[\Ga(h_0)\setminus\seqieq{\{C\}}
          \cup \Ga(h_1[\iota])\setminus\seqieq{\{\neg C\}}\Big]
           \setminus\seqieq{\De_\iota(\IS)}  \Big) \\
 &\subseteq
   \Ga(h_0)\setminus\seqieq{\{C\}} \cup \Big(
     \Big[\De(\IS)\cup \bigcup_{\iota<|\IS|}
       \big( \Ga(h_1[\iota])\setminus\seqieq{\De_\iota(\IS)} \big) \Big]
     \setminus\seqieq{\{\neg C\}}\Big) \\
 &=
   \Ga(h_0)\setminus\seqieq{\{C\}} \ \cup\  
     *(h_1)
     \setminus\seqieq{\{\neg C\}} \\
 &\stackrel{i.h.}{\subseteq}
   \Ga(h_0)\setminus\seqieq{\{C\}} \ \cup\ 
     \seqieq{\Ga(h_1)}
     \setminus\seqieq{\{\neg C\}} 
 \quad\subseteq\quad \seqieq{\Ga(h)}
\end{align*}

\smallskip

\noi
\textbf{Case 2.}
Otherwise, we compute
\begin{align*}
*(h)
 &= \Ga(\I^k_C h_0)\setminus\seqieq{\{C[k]\}} \ \cup\ 
        \Ga(\R_C h_0 h_1[0])\setminus\seqieq{\{\neg C[k]\}} \\
 &= \Big(\{C[k]\}\cup\big(\Ga(h_0)\setminus\seqieq{\{C\}}
     \Big)\setminus\seqieq{\{C[k]\}} \\
 &\quad \cup
    \Big(\Ga(h_0)\setminus\seqieq{\{C\}}
          \cup \Ga(h_1[0])\setminus\seqieq{\{\neg C\}}
     \Big)\setminus\seqieq{\{\neg C[k]\}} \\
 &\subseteq
    \Ga(h_0)\setminus\seqieq{\{C\}} \ \cup\ 
     \Big( \Ga(h_1[0])\setminus\seqieq{\{\neg C[k]\}} \Big)
       \setminus\seqieq{\{\neg C\}} \\
 &\subseteq
    \Ga(h_0)\setminus\seqieq{\{C\}} \ \cup\ 
       *(h_1)
       \setminus\seqieq{\{\neg C\}} \\
 &\stackrel{i.h.}{\subseteq}
   \Ga(h_0)\setminus\seqieq{\{C\}} \ \cup\ 
     \seqieq{\Ga(h_1)}
     \setminus\seqieq{\{\neg C\}}
 \quad\subseteq\quad \seqieq{\Ga(h)}
\end{align*}

\item
$h=\E h_0$:
Then $\Ga(h):=\Ga(h_0)$.

\smallskip

\noi
\textbf{Case 1.}
$\tp(h_0)=\Cut_{C}$:
Assume $\tp(C)\in\{\verum,\tbw\}$, then
\begin{align*}
*(h)
 &= \Ga(\R_C \E h_0\E h_1) \\
 &= \Ga(\E h_0[0])\setminus\seqieq{\{C\}}  \ \cup\ 
    \Ga(\E h_0[1])\setminus\seqieq{\{\neg C\}} \\
 &= \Ga(h_0[0])\setminus\seqieq{\{C\}}  \ \cup\ 
    \Ga(h_0[1])\setminus\seqieq{\{\neg C\}} \\
 &= *(h_0)
 \quad\stackrel{i.h.}{\subseteq}\quad \seqieq{\Ga(h_0)}
 \quad\subseteq\quad \seqieq{\Ga(h)}
\end{align*}
The case that $\tp(C)\notin\{\verum,\tbw\}$ runs similar.

\smallskip

\noi
\textbf{Case 2.}
Otherwise, we compute
\begin{align*}
*(h)
 &= \De(\tp(h_0))\cup\bigcup_{\iota<|\tp(h_0)|}
     \Big(\Ga(\E h_0[\iota])\setminus\seqieq{\De_\iota(\tp(h_0))}\Big) \\
 &= \De(\tp(h_0))\cup\bigcup_{\iota<|\tp(h_0)|}
     \Big(\Ga(h_0[\iota])\setminus\seqieq{\De_\iota(\tp(h_0))}\Big) \\
 &= *(h_0) 
 \quad\stackrel{i.h.}{\subseteq}\quad \seqieq{\Ga(h_0)}
 \quad=\quad \seqieq{\Ga(h)}
\end{align*}
\end{itemize}
\end{proof}

\begin{REM}
For the computation of $\Ga$, the cut-elimination operators $\I^k_C$,
$\R_C$ and $\E$ behave like the following inference symbols:
\mbox{}
\begin{center}
\begin{tabular}{@{}l@{\qquad}l@{\qquad}l@{}}
\AxiomC{$C$}
\LeftLabel{$(\I^k_C)$\quad}
\UnaryInfC{$C[k]$}
\DisplayProof,
&
\AxiomC{$C$}
\AxiomC{$\neg C$}%
\LeftLabel{$(\R_C)$\quad}
\BinaryInfC{$\emptyset$}
\DisplayProof,
&
\AxiomC{$\emptyset$}
\LeftLabel{$(\E)$\quad}
\UnaryInfC{$\emptyset$}
\DisplayProof.
\end{tabular}
\end{center}
\end{REM}

\begin{DEF}
Let $\ceDerN$ be the notation system for cut-elimination on $\DerN$.
The \emph{interpretation $\eval{h}$} is extended inductively 
from $\DerN$ to $\ceDerN$ by defining
\begin{align*}
  \eval{\I^k_C h} &= \opI^k_C(\eval{h})  \\
  \eval{\R_C h_0h_1} &= \opR_C(\eval{h_0},\eval{h_1})  \\
  \eval{\E h} &= \opE(\eval{h})  .
\end{align*}
\end{DEF}

\begin{prop}
For $h\in\ceDerN$ we have
\begin{align*}
  \last(\eval{h}) &= \tp(h)  \\
  \eval{h}(\iota) &= \eval{h[\iota]}  \quad\text{ for } \iota<|\tp(h)| \\
  \Ccrk(\eval{h}) &\le \Ccrk(h)
\end{align*}
\end{prop}

\begin{proof}
By induction on the build-up of $h\in\ceDerN$.
If $h\in\DerN$ then the assertion is inherited from $\DerN$ and
Observation~\ref{observ:formal-semiformal}.
The remaining cases follow from Theorems~\ref{th:I-operator},
\ref{th:R-operator} and~\ref{th:E-operator}.
\end{proof}

\end{section}

\begin{section}{An Abstract Notion of Notation}

We are now interested in studying the size needed by the notations for
sub-derivations of derivations obtained by the cut-elimination
operator. To avoid losing the simple idea in a blurb of notation, we
abstract our problem to a simple term-rewriting system.

\begin{DEF}
An \emph{abstract system of proof notations} is a set $\D$ of
``derivations'', together with two functions
$\sz\cdot,\oo\cdot\colon\D\to\NN\setminus\{0\}$, 
called ``size'' and ``height'', and
a relation $\to\,\subseteq\D\times\D$ called ``reduction to a
sub-derivation'', such that
$d\to d'$ implies $\oo{d'} < \oo{d}$.
\end{DEF}

\begin{observation}[and Definition]\label{obs:abstraction-of-notation-system}
Let $\ForN$ be a notation system for formulae and
$\PrSysForN$ the semiformal proof system over $\ForN$.
A notation system $\DerN=(\DerN,\tp,\cdot[\cdot],\ord,\sz{\cdot})$
for $\PrSysForN$ 
gives rise to 
an abstract system of proof notations
by letting $\D=\DerN$ and defining
$d\to d'$ iff there exists an $n<|\tp(d)|$ with $d'=d[n]$.
\end{observation}

\begin{DEF}
If $\D$ is 
an abstract system of proof notations, then $\Ecl\D$, the
``cut elimination closure'', is the
abstract notation system extending $\D$ 
that is inductively defined by
$$\begin{array}{cccc}
\Rule{d\in\D}{d\in\Ecl\D} &
\Rule{d\in\Ecl\D}{\I d\in\Ecl\D} &
\RuleB{d\in\Ecl\D}{e\in\Ecl\D}{\R de\in\Ecl\D} &
\Rule{d\in\Ecl\D}{\E d\in\Ecl\D} \\
\strut\\
&\sz{\I d} = \sz d + 1 & \sz{\R de} = \sz d + \sz e + 1 & 
\sz{\E d} = \sz d +1\\
\strut\\
\Rule{d\to d'\text{ in $\D$}}{d\to d'}&\Rule{d\to d'}{\I d\to \I d'} & \Rule{e\to e'}{\R de\to\R de'}&
\Rule{d\to d'}{\E d \to \E d'}\\
\strut\\
&&\RuleN{\R de\to\I d}&\RuleB{d\to d'}{d\to d''}{\E d\to \R(\E d')(\E
  d'')}\\
\strut\\
&\oo{\I d} = \oo d & \oo {\R de} = \oo{d} + \oo{e} & \oo{E d} = 2^{\oo
  d}-1
\end{array}
$$
where $\E$, $\R$, $\I$ are new symbols.
\end{DEF}

\begin{proof}
We have to show that whenever $d\to d'$ for $d,d'\in\Ecl\D$ then $\oo d >
\oo{d'}$. We show this by Induction following the inductive definition
of the $\to$ relation in $\Ecl\D$. If $d\to d'$ holds in $\Ecl\D$ because it
already holds in $\D$ then $\oo d>\oo{d'}$ is inherited from $\D$.
The cases $\I d\to \I d'$, $\R d e\to \R d e'$ and $\E
d\to \E d'$ are immediate by induction hypothesis.

For the remaining cases we argue as follows. 
In case $\R de\to \I d$ we calculate
$\oo{\R de}=\oo d + \oo e > \oo d = \oo{\I d}$, since $\oo e > 0$.

In the case $\E d\to \R(E d')(E d'')$ thanks to $d\to d'$ and $d\to d''$
we have $\oo d\geq \oo{d'}+1$ and $\oo d\geq \oo{d''}+1$. 
So, we calculate $\oo{\R (\E
  d')(\E d'')}= \oo{\E d'} + \oo{\E d''} = 2^{\oo{d'}} -1 +
2^{\oo{d''}} -1 < 2^{\oo{d'}} + 2^{\oo{d''}} -1 \leq 2^{\oo d} -1 =
\oo{\E d}$.
\end{proof}

Let $\ForN$ be a notation system for formulae,
$\PrSysForN$ the semiformal proof system over $\ForN$,
$\DerN$ a notation system for $\PrSysForN$, 
$\ceDerN$ the notation system for cut-elimination on $\DerN$
with denoted height $\ord$ and size $\sz{\cdot}$, and
let $\D$ be the 
abstract system of proof notations
associated with $\DerN$ according to 
Observation~\ref{obs:abstraction-of-notation-system}.

\begin{DEF}\label{def:abstraction}
The abstraction $\abstr{h}$ of $h\in\ceDerN$ is obtained by dropping all 
sub- and superscripts.
It can be defined by induction on the build-up of $h\in\ceDerN$:
\begin{itemize}
\item
$h\in\DerN  \Imp  \abstr h := h$,
\item
$h=\I^k_C h_0  \Imp  \abstr h := \I \abstr{h_0}$,
\item
$h=\R_C h_0 h_1  \Imp  \abstr h := \R\, \abstr{h_0}\, \abstr{h_1}$,
\item
$h=\E h_0 \Imp  \abstr h := \E \abstr{h_0}$. 
\end{itemize}
We denote the set of abstractions for $h\in\ceDerN$ by $\abstr\ceDerN$.
\end{DEF}

\begin{observation}[and Definition]\label{obs:abstraction}
The set of abstractions $\abstr\ceDerN$ for $\ceDerN$ is a subsystem
of the cut-elimination closure $\Ecl\DerN$ of $\DerN$ in the following sense:
Let $\to$ denote the reduction to sub-derivation relation of $\Ecl\DerN$,
and define a reduction to sub-derivation relation $\leadsto$ of $\abstr\ceDerN$
in the obvious way by
$\abstr{h}\leadsto \abstr{h'}$ iff there exists an $n<|\tp(h)|$ with $h'=h[n]$.
Then $\abstr\ceDerN=\Ecl\DerN$ and 
$\leadsto\subseteq\to$.
\end{observation}
\end{section}

\begin{section}{Size Bounds}\label{sec:bounds}

We now prove a bound on the size of (abstract) notations for
cut-elimination. By induction on the build up of $\Ecl\D$ we assign
every element a measure that bounds the size of all derivations
reachable from it via iterated use of the $\to$-relation.
A small problem arises in the base case; if $d\to d'$ in $\Ecl\D$
because this holds in $\D$ we have no means of bounding $\sz{d'}$ in
terms of $\sz d$. So we use the usual
trick~\cite{AehligSchwichtenberg00} when a global measure is needed
and assign each element $d$ of $\Ecl\D$ not a natural number but a
monotone function $\szf d$ such that $\sz{d'}\leq \szf d(s)$ for all
$d\to^\ast d'$ whenever $s\in\NN$ is a global bound on the size of all
elements in $\D$.

\begin{DEF}
An abstract system $\D$ of proof notations is called $s$-bounded
(for $s\in\NN$), if for all $d\in\D$ it is the case that 
$\sz d\leq s$.
\end{DEF}

\begin{DEF}
If $\D$ is an abstract system of proof notations and $d\in\D$, then
by $\rstr \D d$ we denote the set $\rstr\D d=\{ d'\mid d\to^\ast
d'\}\subset\D$ considered an abstract system of proof notation with
the structure induced by $\D$. Here $\to^\ast$ denotes the reflexive
transitive closure of $\to$.
\end{DEF}

\begin{DEF}
For $\D$ an abstract system of proof notations and $d\in\D$ we say
that $d$ is $s$-bounded if $\rstr\D d$ is.
\end{DEF}

\begin{DEF} By $\Sz$ we denote the set of all monotone functions from
  $\NN$ to $\NN$.
\end{DEF}

\begin{DEF}
For $\D$ an abstract system of proof notations we define,
a ``size function'' $\szf d\in\Sz$ for every $d\in\Ecl\D$
by induction on the inductive definition of $\Ecl\D$ as follows.
\begin{itemize}
\item For $d\in\D$ we set $\szf d(s)=s$.
\item $\szf{\I d}(s)=\szf d(s)+1$
\item $\szf{\R de}(s)=\max\{\sz d{+}1{+}\szf e(s)\;,\;\szf d(s){+}1\}$
\item $\szf{\E d}(s)=\oo d (\szf d(s) +2)$
\end{itemize}
\end{DEF}
\begin{proof}
The monotonicity of the defined function $\szf d$ is immediately seen from the
definition and the induction hypothesis.
\end{proof}

\begin{prop}\label{prop:szf-bounds-sz}
If $\D$ is 
$s$-bounded then for every $d\in\Ecl\D$ we have $\sz d\leq
\szf d(s)$.
\end{prop}
\begin{proof}
By induction on the inductive definition of $\Ecl\D$.

If $d\in\D$ then $\szf d(s) = s \geq \sz d$, since $\D$ is
$s$-bounded. We calculate $\szf{\I d}(s) = \szf d (s) +1 \geq\sz d +1
=\sz{\I d}$, where we used that $\szf d(s)\geq \sz d$ by induction
hypothesis. Also, $\szf{\R de}(s) \geq \sz d + 1 +\szf e(s)\geq 1+\sz
d + \sz e =\sz{\R de}$, using the induction hypothesis for
$e$. Finally, $\szf{\E d}(s) = \oo d(\szf d(s) +2 ) \geq \szf d(s) +1
\geq \sz d +1 =\sz{\E d}$, where for the first inequality we used that
$\oo d\geq 1$,
and for the second inequality we used the induction hypothesis.
\end{proof}

\begin{Th}\label{th:main}
If $\D$ is $s$-bounded, $d\in\Ecl\D$ and $d\to d'$, then $\szf
d(s)\geq \szf{d'}(s)$.
\end{Th}
\begin{proof}
Induction on the inductive definition of the relation $d\to d'$ in
$\Ecl\D$.

If $d\to d'$ because it holds in $\D$ then $\szf d(s) = s =
\szf{d'}(s)$.

If $\I d\to \I d'$ thanks to $d\to d'$ then $\szf{\I d}(s) = \szf d(s) +1
\geq \szf{d'}(s) + 1 =\szf{\I d'}(s)$, where the inequality is due to
the induction hypothesis.

If $\E d\to\R(\E d')(\E d'')$ thanks to $d\to d'$ and $d\to d''$ we
argue as follows 
$$\begin{array}{cl}
\multicolumn{2}{l}{\szf{\R(\E d')(\E d'')}(s)} \\
=&\max\{\sz{\E d'} {+} 1 {+} \szf{\E d''}(s)\;,\;\szf{\E d'}(s){+}1\}\\
=& \max\{\sz{d'} {+} 2 {+} \oo{d''}(\szf{d''}(s){+}2)\;,\; \oo{d'}(\szf{d'}(s)+2)\}\\
\leq& \max\{\szf{d'}(s) {+} 2 {+} \oo{d''}(\szf{d''}(s){+}2)\;,\; \oo{d'}(\szf{d'}(s)+2)\}\\
\leq& \max\{\szf{d}(s) {+} 2 {+} \oo{d''}(\szf{d}(s){+}2)\;,\; \oo{d'}(\szf{d}(s)+2)\}\\
\leq& \max\{\szf{d}(s) {+} 2 {+} (\oo{d}-1)(\szf{d}(s){+}2)\;,\; (\oo{d}-1)(\szf{d}(s)+2)\}\\
=& \szf{d}(s) {+} 2 {+} (\oo{d}-1)(\szf{d}(s){+}2)\\
=& \oo{d}(\szf d(s){+}2)\\
=& \szf{\E d}(s)
\end{array}
$$
where for the first inequality we used
Proposition~\ref{prop:szf-bounds-sz}, for the second the induction
hypothesis, for the third that, since $d\to d'$ and $d\to d''$, both
$\oo{d'}$ and $\oo{d''}$ are bounded by $\oo d -1$.

If $\E d\to\E d'$ thanks to $d\to d'$ then $\szf{\E d'}(s)
=\oo{d'}(\szf{d'}(s)+2)\leq \oo d(\szf{d'}(s)+2)\leq \oo d(\szf
d(s)+2)=\szf{\E d}(s)$.

If $\R de\to\R de'$ thanks to $e\to e'$, then
$$\begin{array}{cl}
\multicolumn{2}{l}{\szf{\R d e'}(s)} \\
=&\max\{\sz d {+} 1 {+}\szf{e'}(s)\;,\;\szf d(s) {+}1 \}\\
\leq&\max\{\sz d {+} 1 {+}\szf{e}(s)\;,\;\szf d(s) {+}1 \}\\
=&\szf{\R de}
\end{array}
$$
where for the inequality we used the induction hypothesis.

If $\R de\to\I d$ then $\szf{\R de}(s)\geq \szf d(s) +1 =\szf{\I d}(s)$. 
\end{proof}

Now we draw the desired consequences of our main theorem by putting things
together.

\begin{lem}\label{lem:bd-tr-cl}
If $\D$ is 
$s$-bounded, and $d\in\Ecl\D$ then
$\rstr{\Ecl\D} d$ is $\szf d(s)$-bounded.
\end{lem}
\begin{proof}
We first show by induction on the inductive definition of the reflexive transitive
closure that for every $d'\in\rstr{\Ecl\D} d =\{d'\in\Ecl\D\mid
d\to^\ast d'\}$ we have $\szf d (s)\geq \szf {d'}(s)$. The case $d=d'$
is trivial and if $d\to^\ast d'\to d''$ then $\szf
d(s)\geq\szf{d'}(s)$ by induction hypothesis and $\szf{d'}(s)\geq
\szf{d''}(s)$ by Theorem~\ref{th:main}.

Now, by Proposition~\ref{prop:szf-bounds-sz} we know that
$\szf{d'}(s)\geq|d'|$ for $d'\in\Ecl\D$. So, with the previous claim,
for $d'\in\rstr{\Ecl\D}d$ we get $\szf d(s)\geq\szf{d'}(s)\geq |d'|$,
which is the claim.
\end{proof}

\begin{cor}\label{cor:size-bound-Ed-EEd}
If $d\in\D$ is $s$-bounded then $\E d$ is $\oo{d}(s+2)$-bounded and $\E\E
d$ is $2^{\oo{d}}\cdot \oo{d}\cdot (s+4)$-bounded.
\end{cor}
\begin{proof}
Let $d\in\D$ be $s$-bounded and $h:=\oo{d}$. 
First we observe that
$\rstr{\Ecl{(\rstr\D d)}}{d'}=\rstr{\Ecl\D}{d'}$ for any
$d'\in\Ecl{(\rstr\D d)}$. So we can assume without loss of generality
that $\D$ is $s,h$-bounded.

Lemma~\ref{lem:bd-tr-cl} now gives us that $\E d$ is $\szf{\E
 d}(s)$-bounded and $\E\E d$ is $\szf{\E\E d}(s)$-bounded.
We calculate $\szf{\E d} = \oo{d}(\szf d(s)+2) = \oo{d}(s+2)\leq
 h(s+2)$ and $\szf{\E\E d} = \oo{\E d}(\szf{\E d}(d)+2) =
\oo{E d}(h(s+2)+2)\leq (2^h-1)(h(s+2)+2)\leq 2^h\cdot h \cdot (s+4)$.
\end{proof}

Even though the above Corollary covers all the case usually needed in
practise, it is interesting to consider the general case. Recall that
iterated exponentiation $2_n(x)$ is defined inductively by setting
$2_0(x)=x$ and $2_{n+1}(x)=2^{2_n(x)}$. An easy induction shows that
the height $\oo{E^n d}$ of the $n$-times cut-reduced derivation $d$ is
bounded by $2_n(d)$.

\begin{lem}\label{lem:2-k-1-bound-for-E-k-d}
$\szf{E^nd}(s)\leq 2_{n-1}(2\cdot \oo d)\cdot s$ for all $n\geq 1$,
$s\ge2$ and $\oo d\ge2$.
\end{lem}
\begin{proof}
Induction on $n$. 
For the case $n=1$ we compute
$\szf{E d}(s)=\oo d (s+2)\le 2\oo d s$.

For $n=2$ we compute 
$\szf{EE d}(s)=(2^{\oo d}-1)(\oo d (s+2)+2)$.
For $\oo d=2$ and $\oo d=3$ we directly compute that this is bounded
by $2^{2\oo d}s$.
For $\oo d\ge 4$ we compute
$\szf{EE d}(s)\le 2^{\oo d}4\oo d s\le 2^{2\oo d}s$.

Now assume that the claim holds for $n\ge2$.
We then compute 
$\szf{EE^n d}(s)=
\oo{E^nd}(\szf{E^n d}(s)+2)
\leq 2_{n-1}(2^{\oo d}-1)\cdot (2_{n-1}(2\cdot \oo d)\cdot s +2)
\leq 2_{n-1}(2^{\oo d}-1)\cdot 2\cdot 2_{n}(\oo d)\cdot s
\leq 2_n(\oo d)\cdot 2_{n}(\oo d)\cdot s
\leq 2_n(2\cdot\oo d)\cdot s
$
\end{proof}

As an immediate Corollary we obtain
\begin{cor}\label{cor:2-k-1-bound-for-E-k-d}
If $d\in\D$ is $s$-bounded of height $\oo d=h$ for $s\ge2$ and $h\ge2$, 
then $E^k(d)$ is
$2_{k-1}(2\cdot h)\cdot s$-bounded for all $k\geq 1$.
\end{cor}

In Corollary~\ref{cor:2-k-1-bound-for-E-k-d} one should note that the
tower of exponentiations has height only $k-1$. Hence there is one
exponentiation less than the height of the denoted proof.

\bigskip

We conclude this section by remarking that the cut-elimination operator
can be viewed as a polynomial time computable operation.
Assume we modify the size function on $\Ecl\D$ to $\vartheta_k$ by
changing all $\vartheta$ to $\vartheta_k$ and defining for the last case
\begin{itemize}
\item $\szfk{\E d}(s)=(k+1)\cdot (\szf d(s) +2)$
\end{itemize}
Then we obtain as before for $\D$ $s$-bounded, $d\in\Ecl\D$ and $k\in\NN$, 
that $|d|\le\szfk{d}(s)$,
and $d\to d'$ implies $\szfe{k+1}{d}(s)\ge\szfk{d}(s)$.
Hence, for $d\in\D$, $\D$ $s$-bounded, and $\E d\to^k d'$,
we obtain $|d'|\le\szfk{\E d}(s)\le(k+1)\cdot(s+2)$.
From this we can conclude the following observation:
Let $f[i_1,\dots,i_k] := f[i_1]\dots[i_k]$.

\begin{observation}\label{observation:feasibleCutReduction}
The cut-reduction operator for infinitary propositional logic 
is a polynomial time operation in the following sense.

Let $\ForN$ and $\DerN$ be some notation systems for infinitary formulae and
the semiformal system $\PrSysForN$.
Assume that $\ForN$ and $\DerN$ are polynomial time computable,
and that in addition also the functions
\begin{align*}
\ForN\times\NN^{<\om}&\to\ForN\\
A,(i_1,\dots,i_k) &\mapsto A[i_1,\dots,i_k]
\end{align*}
and
\begin{align*}
\DerN\times\NN^{<\om}&\to\DerN\\
h,(i_1,\dots,i_k) &\mapsto h[i_1,\dots,i_k]
\end{align*}
are polynomial time computable.

Then, \ceDerN and the function
\begin{align*}
\DerN\times\NN^{<\om}&\to\ceDerN\\
h,(i_1,\dots,i_k) &\mapsto (\E h)[i_1,\dots,i_k]
\end{align*}
are polynomial time computable.
\end{observation}

\end{section}

\begin{section}{Bounded Arithmetic}\label{sec:bounded-arithmetic}

Our proof-theoretic investigations are very much independent of
the exact choice of the language.
Therefore, we will be very liberal and allow symbols for all ptime 
functions.

\begin{DEF}[Language of Bounded Arithmetic]
\emph{The language $\lba$ of Bounded Arithmetic}  contains as non-logical symbols
$\{ =, \le \}$ for the binary relation ``equality'' and ``less than or equal'',
and a symbol for each ptime function.
In particular, it includes
a constant $c_a$ for $a\in\NN$ whose interpretation in the standard model 
$\NN$ is $c_a^\NN=a$,
unary function symbols $|\cdot|$ and $\zhl{\cdot}$ which have their
standard interpretation given by $(|c_a|)^\NN=n$ and $(\zhl{c_a})^\NN=2^n$
where $n$ is the length of the binary representation of $a$,
and the binary function symbols $\min$ and $\sma$ 
whose standard interpretation are minimisation and
$(c_a\sma c_b)^\NN=2^{n\cdot m}$ where $n$ and $m$ are the lengths of the
binary representations of $a$ resp.\ $b$.
We will often write $\ul n$ instead of $c_n$, and $0$ for $c_0$.

\emph{Atomic formulae} are of the form $s=t$ or $s\le t$ where $s$ and $t$ 
are terms.
\emph{Literals} are expressions of the form $A$ or $\neg A$ 
where $A$ is an atomic formula.
Formulas are build up from literals by means of $\und$, $\oder$, 
$(\forall x)$, $(\exists x)$.
The \emph{negation $\neg C$ for a formula $C$} is defined via de Morgan's laws.
Negation extends to sets of formulae in the usual way by applying it to 
their members individually.

Let $\cC$ be a set of $\lba$-formulae (think of $\sib i$),
and $A$ an $\lba$-formula.
We define the \emph{$\cC$-rank of $A$,} denoted $\Crk(A)$,
by induction on the build-up of $A$:
\begin{itemize}
\item
If $A\in\cC\cup\neg\cC$, let $\Crk(A):=0$.
\item
If $A=B\land C$ or $A=B\lor C$, let 
$\Crk(A):=1+\max\{\Crk(B),\Crk(C)\}$.
\item
If $A=(\forall x)B$ or $A=(\exists x)B$, let
$\Crk(A):=1+\Crk(B)$.
\end{itemize}
\end{DEF}

We will use the following standard abbreviations.
\begin{DEF}[Abbreviations]
The expression $A\imp B$ denotes the expression $\neg A \oder B$.
The expression $s<t$ denotes $\neg t\le s$.
Bounded quantifiers are introduced as follows:
$(\forall x\klg t)A$ denotes $(\forall x)A_x(\min(x,t))$,
$(\exists x\klg t)A$ denotes $(\exists x)A_x(\min(x,t))$,
$(\forall x\kl t)A$ denotes $(\forall x\klg t)(x<t\imp A)$,
$(\exists x\kl t)A$ denotes $(\exists x\klg t)(x<t\und A)$,
where $x$ may not occur in $t$.
\end{DEF}

\begin{DEF}[Bounded Formulas]
The set $\FOR$ of bounded \lba-formulae is the set of \lba-formulae
consisting of literals and closed under 
$\und$, $\oder$, $(\forall x\klg t)$, $(\exists x\klg t)$.
\end{DEF}

We now define a restricted (also called ``strict'') delineation of 
bounded formulae.
\begin{DEF}
The set \ssib{d} is the subset of bounded \lba-formulae whose elements
are of the form
\[
(\exists x_1\klg t_1) (\forall x_2\klg t_2) \dots 
  (Q x_d\klg t_d) (\bar Q x_{d+1}\klg |t_{d+1}|) A(\vec x)
\]
with $Q$ and $\bar Q$ being of the corresponding alternating quantifier 
shape, and  $A$ being quantifier free.
\end{DEF}

\begin{DEF}
As axioms we allow all disjunctions of literals, i.e., all
disjunctions $A$ of literals such that $A$ is true in $\NN$ under 
any assignment.
Let us denote this set of axioms by \basic.
\end{DEF}

We will base the definition of Bounded Arithmetic theories on
a somewhat stronger normal form of induction.
Let $|\cdot|_m$ denote the $m$-fold iteration of the function symbol 
$|\cdot|$.
\begin{DEF}
Let $\Ind(A,z,t)$ denote the expression
\[
  A_z(0)\und(\forall z\kl t)(A\imp A_z(\suc z))  \imp  A_z(t) \enspace.
\]
The set $\Phi\dilind{m}$ consists of all expressions of the form
\[
 \Ind(A,z,\zhl{|t|_m})
\]
with $A\in\Phi$, $z$ a variable and $t$ an \lba-term.
\end{DEF}

This restricted form of induction implies the usual form, because the 
following can be proven from \basic alone.
\[
  \Ind(A(\min(t,z)),z,\zhl{t}) \imp \Ind(A(z),z,t)
\]

\end{section}

\begin{section}{Notation system for Bounded Arithmetic formulae}

Let \ForNBA be the set of closed formulae in \FOR.
We define the outermost connective function on \ForNBA by
\[
  \tp(A) := \begin{cases}
    \verum & A \text{ true literal} \\
    \falsum & A \text{ false literal} \\
    \tbw & A \text{ is of the form } A_0\und A_1 
                \text{ or } (\forall x)B \\
    \tbv & A \text{ is of the form } A_0\oder A_1 
                \text{ or } (\exists x)B  \enspace,
  \end{cases}
\]
and the sub-formula function on $\ForNBA\times\NN$ by
\[
  A[n] := \begin{cases}
    A & A \text{ literal} \\
    A_{\min(n,1)} & A \text{ is of the form } A_0\und A_1 
                \text{ or } A_0\oder A_1 \\
    B_x(\ul n) & A \text{ is of the form } (\forall x)B
                \text{ or } (\exists x)B  \enspace.
  \end{cases}
\]
The rank and negation functions for the notation system
are those defined for \lba.

We didn't have much choice on how to render $\FOR$ into a notation
system for formulae. Nevertheless, the above definition already shows
that we have to work with a non-trivial intensional equality. The
reason is that, even though in the process of the propositional
translation we can make sure that we only have closed formulae, this
still is not enough; we do have other closed terms than just the
canonical ones.

Consider, for example, an arithmetical derivation ending in
$$
\AxiomC{$\vdots$}\noLine
\UnaryInfC{$B(f(\ul 0))$}
\UnaryInfC{$\exists x.B(x)$}
\DisplayProof
$$
where $f$ is some function symbol. In the propositional translation we
have to provide some witness $i$ for the $\tbv_{\exists x.B(x)}^i$-inference.
The ``obvious'' choice seems to take $i=f^\NN(0)$. But this would
require a derivation of $(\exists x.B(x))[f^\NN(0)]=B(\ul{f^\NN
 (0)})$. The translation of the sub-derivation, on the other hand, gives us a
derivation of $B(f(\ul 0))$. So, in order to make this a correct
inference in the propositional translation, he have to consider
$B(f(\ul 0))$ and $B(\ul{f^\NN(0)})$ as intensionally equal. Note that
both formulae are extensionally equal.

We will now define an intensional equality which provides the above
described identification.
For $t$ a closed term its numerical value $t^\NN\in\NN$ is defined in
the obvious way.
Let $\red$ denote the rewriting relation obtained from
\[
  \setof{(t,\ul{t^\NN})}{t \text{ a closed term}} \enspace.
\]
For example, 
\[ (\forall x)(x\le\shr{(\ul 5\cdot\ul 3)})  \red
   (\forall x)(x\le\ul 7) \enspace. \]
Let $\eqnat$ denote the reflexive, symmetric and transitive closure
of $\red$.

\begin{prop}
The just defined system consisting of 
\ForNBA, $\tp$, $\cdot[\cdot]$, $\neg$, $\rk$ and $\eqnat$ 
forms a notation system for formulae 
in the sense of Definition~\ref{def:NotationSystemForFormulas}.
\end{prop}

\begin{REM}
It is an open problem what the complexity of $\eqnat$ is
(assuming a usual feasible arithmetisation of syntax).
However, if the depth of expressions is restricted, and the number
of function symbols representing polynomial time functions is also
restricted to a finite subset, 
then the relation $\eqnat$ is polynomial time decidable.
I.e., let $\eqnat^k$ denote the restriction of $\eqnat$ to expressions
of depth $\le k$ in which at most the first $k$ function symbols occur.
Then, for each $k$, the relation $\eqnat^k$ is a polynomial time 
predicate.
\end{REM}

From now on, we will assume that \ForNBA implicitly contains such 
a constant $k$ without explicitly mentioning it.
All formulae and terms used in \ForNBA are thus assumed to obey
the abovementioned restriction on occurrences of function symbols
and depth.
We will come back to this restriction at relevant places.
The next observation already makes use of this assumption.

\begin{observation}
All relations and functions in \ForNBA are polynomial time computable.
\end{observation}
\begin{proof}
Under the just fixed convention, the relation
$\eqnat$ is actually $\eqnat^k$ for some $k$.
\end{proof}

\begin{DEF}
Let \bainf denote the semiformal proof system over \ForNBA
according to Definition~\ref{def:semiformal_proof_systems}.
\end{DEF}

\end{section}

\begin{section}{A notation system for \bainf}

\begin{DEF}
The \emph{finitary proof system $\bastar$} is the proof system over 
$\FOR,\eqnat$ which is given by the following set of inference
symbols.
\mbox{}\\[3ex]
\begin{tabular}{@{}ll@{}}
\AxiomC{\mbox{}}
\LeftLabel{$(\Ax_\De)$\quad}
\RightLabel{\quad if $\bigvee\De\in\basic$}
\UnaryInfC{$\De$}
\DisplayProof
\\[2ex]
\AxiomC{$A_0$}
\AxiomC{$A_1$}
\LeftLabel{$(\tbw_{A_0\wedge A_1})$\quad}
\BinaryInfC{$A_0\wedge A_1$}
\DisplayProof
&
\AxiomC{$A_k$}
\LeftLabel{$(\tbv^k_{A_0\vee A_1})$\quad}
\RightLabel{\quad $(k\in\{0,1\})$}
\UnaryInfC{$A_0\vee A_1$}
\DisplayProof
\\[3ex]
\AxiomC{$A_x(y)$}
\LeftLabel{$(\tbw^y_{(\forall x) A})$\quad}
\UnaryInfC{$(\forall x) A$}
\DisplayProof
&
\AxiomC{$A_x(t)$}
\LeftLabel{$(\tbv^{t}_{(\exists x) A})$\quad}
\UnaryInfC{$(\exists x) A$}
\DisplayProof
\\[3ex]
\AxiomC{$\neg F, F_y(\suc y)$}
\LeftLabel{$(\ind^{y,t}_{F})$\quad}
\UnaryInfC{$\neg F_y(0), F_y(\zhl{t})$}
\DisplayProof
&
\AxiomC{$\neg F, F_y(\suc y)$}
\LeftLabel{$(\ind^{y,n,i}_{F})$\quad}
\RightLabel{\quad $(n,i\in\NN)$}
\UnaryInfC{$\neg F_y(\ul n), F_y(\ul{n+2^i})$}
\DisplayProof
\\[3ex]
\AxiomC{$C$}
\AxiomC{$\neg C$}%
\LeftLabel{$(\Cut_C)$\quad}
\BinaryInfC{$\emptyset$}
\DisplayProof
\end{tabular}
\end{DEF}

According to Definition~\ref{def:quasi-derivations}, 
a $\bastar$-quasi derivation $h$ is equipped with functions
$\Ga(h)$ denoting the endsequent of $h$,
$\hgt(h)$ denoting the height of $h$,
and $\dsz{h}$ denoting the size of $h$.

In our finitary proof system Sch{\"u}tte's $\omega$-rule~\cite{Schuette51}
is replaced by rules
with Eigenvariable conditions. Of course, the precise name of the
Eigenvariable does not matter, as long as it \emph{is} an
Eigenvariable. For this reason, we think of
the inference symbols
$\tbw^y_{(\forall x) A}$, $\ind^{y,t}_{F}$, and $\ind^{y,n,i}_{F}$
in $\bastar$-quasi derivations
as binding the variable $y$ in the respective sub-derivations.
Fortunately, we don't have to make this intuition precise, as we will
always substitute only closed (arithmetical) terms into
$\bastar$-derivations and therefore no renaming of bound variables will be
necessary; hence we don't have to define what this renaming would mean. Note,
however, that the details of Definition~\ref{def:bastarderivation} of
$\bastar$-derivations and Definition~\ref{def:subst-bastar} of
substitution become obvious with this intuition on mind.

\begin{DEF}[Inductive definition of $\varDeriv{\vec x}d$]\label{def:bastarderivation}
For $\vec x$ a finite list of disjoint variables and $d=\IS d_0\dots d_{n-1}$ 
a $\bastar$-quasi-derivation we inductively define the relation
$\varDeriv{\vec x}d$ that $d$ is a $\bastar$-derivation with free
variables among $\vec x$ as follows.
\begin{itemize}
\item
If $\varDeriv{\vec x,y}{h_0}$ and 
$\IS\in\{\tbw^y_{(\forall x) A},\ind^{y,t}_{F},\ind^{y,n,i}_{F}\}$ for
some $A,F,t,n,i$,
and $\fv(\Ga(\IS h_0))\subset\{\vec x\}$
then
$\varDeriv{\vec x}{\IS h_0}$.

\item
If $\varDeriv{\vec x}{h_0}$ and $\fv((\exists x)A),\fv(t)\subseteq\{\vec x\}$ then
$\varDeriv{\vec x}{\tbv^{t}_{(\exists x) A} h_0}$.

\item
If $\varDeriv{\vec x}{h_0}$, $\varDeriv{\vec x}{h_1}$ and
$\fv(C)\subseteq\{\vec x\}$ then
$\varDeriv{\vec x}{\Cut_C h_0h_1}$.

\item
If $\fv(\De)\subseteq\{\vec x\}$ then $\varDeriv{\vec x}{\Ax_\De}$,

\item
If $\varDeriv{\vec x}{h_0}$, $\varDeriv{\vec x}{h_1}$ and
  $\IS= \tbw_{A_0\wedge A_1}$ with $\fv({A_0\wedge A_1})\subset\{\vec x\}$
then
$\varDeriv{\vec x}{\IS h_0h_1}$.

\item
If $\varDeriv{\vec x}{h_0}$ and
  $\IS=\tbv^k_{A_0\vee A_1}$ 
with $\fv({A_0\vee A_1})\subset\{\vec x\}$
then
$\varDeriv{\vec x}{\IS h_0}$.
\end{itemize}

A $\bastar$-derivation is a $\bastar$-quasi derivation $h$ such that for
some $\vec x$ it holds $\varDeriv{\vec x}h$. We call a
$\bastar$-derivation $h$ \emph{closed}, if $\varDeriv\emptyset h$.
\end{DEF}

\begin{prop}
If $\varDeriv{\vec x}h$ then $\fv(\Ga(h))\subseteq\{\vec x\}$. In
particular $\fv(\Ga(h))=\emptyset$ for closed $h$.
\end{prop}
\begin{proof}
Trivial induction on the inductive definition of $\varDeriv{\vec x}h$.
\end{proof}

\begin{DEF}\label{def:subst-bastar}
For $h$ a $\bastar$-derivation, $y$ a variable and $t$ a closed term of
Bounded Arithmetic we define the substitution 
$h(t/y)$ inductively by setting
$(\IS h_0\ldots h_{n-1})(t/y)$ to be $\IS(t/y) h_0(y/t)\ldots
h_{n-1}(t/y)$ if $\IS$ is not of the form
$\tbw^y_{(\forall x) A}$, $\ind^{y,t}_{F}$, or $\ind^{y,n,i}_{F}$ with
the same variable $y$, and $\IS h_0\ldots h_{n-1}$ otherwise.

Substitution for inference symbols is defined by setting
$$\begin{array}{lclclcl}
\Ax_\De(t/y) &=& \Ax_{\De(t/y)} \\
\tbw_{A_0\wedge A_1}(t/y) &=& \tbw_{(A_0\wedge A_1)(t/y)}
&\qquad&
\tbv^k_{A_0\wedge A_1}(t/y) &=& \tbv^k_{(A_0\wedge A_1)(t/y)}
\\
\tbw^z_{(\forall x) A}(t/y) &=& \tbw^z_{((\forall x) A)(t/y)}
&&
\tbv^{t'}_{(\exists x) A}(t/y) &=& \tbv^{t'(t/y)}_{((\exists x) A)(t/y)}
\\
\ind^{z,t'}_{F}(t/y) &=& \ind^{z,t'(t/y)}_{F(t/y)}
&&
\ind^{z,n,i}_{F}(t/y) &=& \ind^{z,n,i}_{F(t/y)}
\end{array}
$$
\end{DEF}

We now show the substitution property for $\bastar$-derivations. The
formulation of Lemma~\ref{lem:bastar-subst} might look a bit strange
with ``$\subseteq$'' instead of the more familiar equality. The reason
is, that a substitution may make formulae equal which are not equal
without the substitution.

Recalling however Definition~\ref{def:turnstile}, we note that
derivations $h$ in fact prove every superset of $\Ga(h)$. Of course,
an easy consequence of Lemma~\ref{lem:bastar-subst} is that if
$\Ga(h)\subset\Delta$ then $\Ga(h(t/y))\subset\Delta(t/y)$.

\begin{lem}\label{lem:bastar-subst}
Assume $\varDeriv{\vec x}h$ and let $y$ be a variable and $t$ a closed term, 
then $\varDeriv{\vec x\setminus\{y\}} h(t/y)$ 
and moreover $\Ga(h(t/y))\subseteq(\Ga(h))(t/y)$.
\end{lem}
\begin{proof}
We argue by induction on the build-up of $h$.

In the cases where no substitution occurs (as $h=\IS\dots$ with $\IS$
of the form
$\tbw^y_{(\forall x) A}$, $\ind^{y,t}_{F}$, or $\ind^{y,n,i}_{F}$ with
the same variable $y$) both claims are trivial.

Otherwise, by induction hypothesis, we know that the sub-derivations
are $\bastar$-derivations with the correct set of free variables;
since substitution is also carried out in the inference symbols, the
$y$ in the variable conditions for $\Cut_C$ and $\tbv^t_{(\exists x)A}$ will
also disappear due to the substitution. The Eigenvariable condition
$z\not\in\fv(\Ga(h))$ will follow once we have shown the second claim.

For the second claim we compute by induction hypothesis
\[
 \Ga((h(t/y))(\iota)) = \Ga((h(\iota))(t/y)) \subseteq \Ga((h(\iota)))(t/y)
\]
Hence
\begin{align*}
\Ga(h(t/y)) &= \De(\last(h(t/y)))\cup\bigcup_{\iota<|\last(h)|} \Big(
    \Ga((h(t/y))(\iota)) \setminus \seqeqnat{\De_\iota(\last(h(t/y)))}
  \Big)\\
 &\stackrel{i.h.}{\subseteq}
  \De(\last(h))(t/y)\cup\bigcup_{\iota<|\last(h)|} \Big(
    \Ga((h)(\iota))(t/y) \setminus \seqeqnat{\De_\iota(\last(h))(t/y)}
  \Big)\\
 &\stackrel{!!!}{\subseteq}
   \Big( \De(\last(h))\cup\bigcup_{\iota<|\last(h)|} \big(
    \Ga((h)(\iota)) \setminus \seqeqnat{\De_\iota(\last(h))}
  \big) \Big)(t/y)\\
 &= \Ga(h)(t/y)
\end{align*} 
This finishes the proof.
\end{proof}

We will now define the ingredients for a notation system
for \bainf, which forms the embedding of \bastar into \bainf.

Let \DerNBA be the set of closed \bastar-derivations.

For each $h\in\DerNBA$ we define the denoted last inference $\tp(h)$ 
as follows:
Let $h=\IS h_0\dots h_{n-1}$,
\[ \tp(h) := \begin{cases}
\Ax_A
  & \text{if } \IS=\Ax_\De,
   \text{ where $A$ is the ``least'' true literal in } \De \\
\tbw_{A_0\und A_1}
  & \text{if } \IS=\tbw_{A_0\und A_1}  \\
\tbv^{k}_{A_0\oder A_1}
  & \text{if } \IS=\tbv^k_{A_0\oder A_1}  \\
\tbw_{(\forall x)A}
  & \text{if } \IS=\tbw^y_{(\forall x)A}  \\
\tbv^{t^\NN}_{(\exists x)A}
  & \text{if } \IS=\tbv^t_{(\exists x)A}  \\
\Rep
  & \text{if } \IS=\ind^{y,t}_{F}  \\
\Rep
  & \text{if } \IS=\ind^{y,n,0}_{F}  \\
\Cut_{F_y(\ul{n+2^i})}
  & \text{if } \IS=\ind^{y,n,i+1}_{F}  \\
\Cut_{C}
  & \text{if } \IS=\Cut_C
\end{cases} \]

For each $h\in\DerNBA$ and $j\in\NN$ we define the denoted sub-derivation
$h[j]$ as follows:
Let $h=\IS h_0\dots h_{n-1}$.
If $j\ge|\tp(h)|$ let $h[j] := \Ax_{0=0}$.
Otherwise, assume $j<|\tp(h)|$ and define
\[ h[j] := \begin{cases}
h_{\min(j,1)}
  & \text{if } \IS=\tbw_{A_0\und A_1}  \\
h_0
  & \text{if } \IS=\tbv^k_{A_0\oder A_1} \\
h_0(\ul j/y)
  & \text{if } \IS=\tbw^y_{(\forall x)A}  \\
h_0
  & \text{if } \IS=\tbv^t_{(\exists x)A}   \\
\ind^{y,0,|t|^\NN}_{F}h_0
  & \text{if } \IS=\ind^{y,t}_{F}  \\
h_0(\ul n/y)
  & \text{if } \IS=\ind^{y,n,0}_{F}  \\
\ind^{y,n,i}_{F}h_0
  & \text{if } \IS=\ind^{y,n,i+1}_{F} \text{ and } j=0   \\
\ind^{y,n+2^i,i}_{F}h_0
  & \text{if } \IS=\ind^{y,n,i+1}_{F} \text{ and } j=1   \\
h_j
  & \text{if } \IS=\Cut_C
\end{cases} \]

The denoted end-sequent function on $\DerNBA$ is given by $\Ga$
computed according to Definition~\ref{def:quasi-derivations}.
The size function $\sz{\cdot}$ on $\DerNBA$ is given by $\sz{h}:=\dsz{h}$.

To define the denoted height function 
we need some analysis yielding an upper
bound to the log of the lengths of inductions 
which may occur during the embedding
(we take the log as this bounds the height of the derivation
tree which embeds the application of induction).
Let us first assume $m$ is such an upper bound, and let us define
the denoted height $\ord_m(h)$ of $h$ relative to $m$:
For a $\bastar$-derivation $h=\cI h_0\dots h_{n-1}$ we define
\[
\ord_m(h) := \begin{cases}
\ord_m(h_0)+i+1            & \text{if }\cI=\ind^{y,n,i}_{F} \\
\ord_m(h_0)+m+1            & \text{if }\cI=\ind^{y,t}_{F} \\
1+\sup_{i<n}\ord_m(h_i)  & \text{otherwise}  
\end{cases}
\]
Observe that $\ord_m(h)>0$ (in particular, $\ord(\Ax_\De)=1$).

To fill the gap of providing a suitable upper bound function of 
\bastar-derivations
we first need to fix monotone bounding terms for any term in \lba.

\subsection*{Bounding terms}

For a term $t$ we define a term $\bd(t)$ which represents a monotone 
function with the following property:
If $\fv(t)=\{\vec x\}$ then
\[  (\forall\vec n)\qquad 
 t_{\vec x}(\vec{\ul n})^\NN \quad\le\quad
   \bd(t)_{\vec x}(\vec{\ul n})^\NN
\]
Let $x_0,x_1,x_2,\dots$ be a fixed list of free variables.
We fix for each function symbol $f$ of arity $n$ a 
\emph{monotone bounding term} $T_f$ with
$\fv(T_f)\subseteq\{x_0,\dots,x_{n-1}\}$.
E.g., assume that we have fixed for each function symbol $f$ in our language
a number $c_f\in\NN$ such that 
$(\forall\vec n)|f^\NN(\vec n)|\le \max\{2,|\vec n|\}^{2^{c_f}}$ holds.
We then can define
\[  T_f \quad:=\quad
  \underbrace{(\max\{2,\vec x\})\sma\dots\sma(\max\{2,\vec x\})
              }_{2^{c_f}\text{ times}}
    \enspace.
\]
As the only exception we demand that $T_{|\cdot|}:=|x_0|$.

Now, let $t$ be a term.
If $t$ is a closed term, let $\bd(t) := \underline{t^\NN}$.
If $t=ft_1\dots t_n$ is not a closed term, let
$\bd(t):=(T_f)_{\vec x}(\bd(t_1),\dots,\bd(t_n))$.

\subsection*{Bounding terms for
   $\bastar$-derivations}

For $h\in\DerNBA$, the bounding term $\bd(h)$ is intended to
bound any variable which occurs during the embedding of $h$, 
and the term $|\ibd(h)|$ is intended to bound the length of any
induction which occurs during the embedding of $h$.

Let $h=\cI h_0\dots h_{n-1}$ be in \DerNBA..
We define
\begin{align*}
\bd(h)&:=
\begin{cases}
\max(\bd(h_0(\ul{\bd(t)}/y)),\bd(t))
& \text{ if } \cI=\tbw^y_{(\forall x\le t)A}  \\
\max(\bd(h_0),\bd(t))
& \text{ if } \cI=\tbv^t_{(\exists x)A}  \\
\max(\bd(h_0(\ul{2^{|\bd(t)|}}/y)),2^{|\bd(t)|})
& \text{ if } \cI=\ind^{y,t}_F  \\
\max(\bd(h_0(\ul{n+2^i}/y)),n+2^i)
& \text{ if } \cI= \ind^{y,n,i}_F  \\
\max(\bd(h_0),\dots,\bd(h_{n-1}))
& \text{ otherwise.}
\end{cases}  \\[2ex]
\ibd(h)&:=
\begin{cases}
\ibd(h_0(\ul{\bd(t)}/y))
& \text{ if } \cI=\tbw^y_{(\forall x\le t)A}  \\
\max(\ibd(h_0(\ul{2^{|\bd(t)|}}/y)),2^{|\bd(t)|})
& \text{ if } \cI=\ind^{y,t}_F  \\
\max(\ibd(h_0(\ul{n+2^i}/y)),2^i)
& \text{ if } \cI= \ind^{y,n,i}_F  \\
\max(\ibd(h_0),\dots,\ibd(h_{n-1}))
& \text{ otherwise.}
\end{cases}  
\end{align*}

Now we can define the denoted height function
$\ord(h):=\ord_{|\ibd(h)|}(h)$ for $h\in\DerNBA$.

\begin{Th}
The just defined system consisting of \DerNBA, $\tp$, $\cdot[\cdot]$, $\Ga$,
$\ord(\cdot)$ and $|\cdot|$ forms a notation system for \bainf
in the sense of Definition~\ref{def:notation-for-proof-system}.
\end{Th}

\begin{proof}
First, we observe that $\ord(\cdot)$ satisfies the following 
monotonicity property:
\begin{equation}\label{Monotonicity}
m\le m' \Imp  \ord_m(h)\le\ord_{m'}(h) \enspace.
\end{equation}
We also observe the following substitution property by inspection:
\begin{equation}\label{Substitution}
\ord_m(h(t/y))=\ord_m(h) \enspace.
\end{equation}

We prove the following slightly more general assertion:
\begin{equation}\label{GeneralAssertion}
m\ge|\ibd(h)| \Und  i<|\tp(h)|  \Imp  \ord_m(h[i])<\ord_m(h)
\end{equation}
Then the assertion of the theorem follows using the monotonicity 
property \eqref{Monotonicity}, as $\ibd(h[i])\le\ibd(h)$.

The proof of \eqref{GeneralAssertion} is by induction on the build-up 
of $h$.
Let $h=\cI h_0\dots h_{n-1}$.

First assume that $h[i]=h_j(t/y)$.
The definition of $\ord_m$ immediately shows that in this case
$\ord_m(h)=1+\sup_{i<n}\ord_m(h_i)$.
The substitution property \eqref{Substitution} shows that 
$\ord_m(h_j(t/y))=\ord_m(h_j)$. 
Hence
\[
\ord_m(h)>\ord_m(h_j)= \ord_m(h_j(y/k)) = \ord_m(h[i])  \enspace.
\]

The remaining cases are the following ones:

If $h=\ind^{y,t}_{F} h_0$, then $h[0]=\ind^{y,0,|t|}_{F} h_0$.
 As $|t|\le|\bd(t)|<|\ibd(h)|\le m$ we obtain
 \[
 \ord_m(h[0])=\ord_m(h_0)+|t|+1<\ord_m(h_0)+m+1=\ord_m(h) \enspace.
 \]

If $h=\ind^{y,n,k+1}_{F} h_0$, 
then $h[i]= \ind^{y,n',k}_{F} h_0$ for some $n'$
 Hence
 \[
 \ord_m(h[i])=\ord_m(h_0)+k+1<\ord_m(h_0)+k+2=\ord_m(h) \enspace.
 \]

Thus, assertion \eqref{GeneralAssertion} is proven.
The Theorem follows using the next Proposition which shows the
local faithfulness property of the denoted end-sequent function $\Ga$.
\end{proof}

\begin{prop}
$\Ga$ satisfies the local faithfulness property:
Let $h\in\bastar$, then
\[  
\De(\tp(h))\cup \bigcup_{\iota<|\tp(h)|} 
        \Big(\Ga(h[\iota])\setminus\seqeqnat{\De_\iota(\tp(h))}) \Big)
\subseteq \seqeqnat{\Ga(h)}
\enspace. \]
\end{prop}

\begin{proof}[Proof by induction on $\ord(h)$]
Let $h=\IS h_0\dots h_{n-1}\in\DerNBA$.
We abbreviate
\[  
*(h) \quad:=\quad
   \De(\tp(h))\cup \bigcup_{\iota<|\tp(h)|} 
        \Big(\Ga(h[\iota])\setminus\seqeqnat{\De_\iota(\tp(h))}) \Big)
\enspace. \]

\medskip

\noi
\textbf{Case 1.}
$\IS=\Ax_\De$: 
Let $A$ be the ``least'' true literal in $\De$,
then 
\[
  *(h)=\De(\Ax_A)=\{A\}\subseteq\De=\Ga(h)
\]

\medskip

\noi
\textbf{Case 2.}
$\IS=\tbw_{C}$ for $C=A_0\land A_1$:
$\tp(h)=\tbw_{C}$, 
$h[0]=h_0$ and $C[0]=A_0$, and
$h[\iota]=h_1$ and $C[\iota]=A_1$ for $\iota>0$,
hence
\begin{align*}
*(h) 
 &= \{A_0\und A_1\}
     \cup ({\Ga({h_0})}\setminus \seqeqnat{\{A_0\}})
     \cup ({\Ga({h_1})}\setminus \seqeqnat{\{A_1\}}) \\
 &= {\Ga(h)}
\end{align*}

\medskip

\noi
\textbf{Case 3.}
$\IS=\tbv^{k}_{A_0\oder A_1}$

\medskip

\noi
\textbf{Case 4.}
$\IS=\tbw^y_{(\forall x)A}$:
$\tp(h)=\tbw_{(\forall x)A}$, and
$h[\iota]=h_0(\ul\iota/y)$ and $((\forall x)A)[\iota]=A(\ul\iota/x)$
for $\iota\in\NN$
hence
\begin{align*}
*(h) 
 &= \{(\forall x)A\}
     \cup \bigcup_{i\in\NN}
       ({\Ga({h_0(\ul i/y)})}\setminus \seqeqnat{\{A(\ul i/x)\}}) \\
 &\subseteq \{(\forall x)A\}
     \cup \bigcup_{i\in\NN}
       ({\Ga({h_0})(\ul i/y)}\setminus \seqeqnat{\{A(\ul i/x)\}}) \\
 &\stackrel{(1)}{=} \{(\forall x)A\}
     \cup \bigcup_{i\in\NN}
       (\Ga({h_0})\setminus \seqeqnat{\{A\}}) \\
 &= \Ga(h)
\end{align*}
$(1)$: uses Eigenvariable condition.

\medskip

\noi
\textbf{Case 5.}
$\IS=\tbv^t_{(\exists x)A}$:
$\tp(h)=\tbv^{t^\NN}_{(\exists x)A}$ and $h[0]=h_0$,
hence
\begin{align*}
*(h) 
 &= \{(\exists x)A\}
     \cup (\Ga({h_0}) \setminus \seqeqnat{\{A(\ul t^\NN/x)\}}) \\
 &= \{(\exists x)A\}
     \cup (\Ga({h_0}) \setminus \seqeqnat{\{A(t/x)\}}) \\
 &= \Ga(h)
\end{align*}

\medskip

\noi
\textbf{Case 6.}
$\IS=\ind^{y,t}_{F}$:
$\tp(h)=\Rep$ and $h[0]=\ind^{y,0,|t^\NN|}_{F}h_0$,
hence
\begin{align*}
*(h) 
 &= \emptyset \cup \Ga({\ind^{y,0,|t^\NN|}_{F}h_0 }) \\
 &= \{\neg F_y(\ul0), F_y(\ul{0+\zhl{t^\NN}}) \}
   \cup (\Ga(h_0) \setminus \seqeqnat{\{\neg F, F_y(\suc y)\}} ) \\
 &\subseteq \seqeqnat{\{\neg F_y(0), F_y(\zhl{t}) \}}
   \cup (\Ga(h_0) \setminus \seqeqnat{\{\neg F, F_y(\suc y)\}} ) \\
 &\subseteq \seqeqnat{\Ga(h)}
\end{align*}

\medskip

\noi
\textbf{Case 7.}
$\IS=\ind^{y,n,0}_{F}$:
$\tp(h)=\Rep$ and $h[0]=h_0(\ul n/y)$,
hence
\begin{align*}
*(h) 
 &= \emptyset \cup \Ga(h_0(\ul n/y)) \\
 &\subseteq \Ga(h_0)(\ul n/y) \\
 &\stackrel{(2)}{\subseteq}
  \seqeqnat{\{\neg F_y(\ul n),F_y(\suc\ul n)\}}
    \cup (\Ga(h_0)\setminus \seqeqnat{\{\neg F,F_y(\suc y)\}})  \\
 &\subseteq  \seqeqnat{\Big(
  \{\neg F_y(\ul n),F_y(\ul{n+1})\}
    \cup (\Ga(h_0)\setminus \seqeqnat{\{\neg F,F_y(\suc y)\}}) \Big)}  \\
 &= \seqeqnat{\Ga(h)}
\end{align*}
$(2)$ uses Eigenvariable condition.

\medskip

\noi
\textbf{Case 8.}
$\IS=\ind^{y,n,i+1}_{F}$:
$\tp(h)=\Cut_{F_y(\ul{n+2^i})}$,
$h[0]=\ind^{y,n,i}_{F}h_0$, and
$h[1]=\ind^{y,n+2^i,i}_{F}h_0$,
hence (abbreviating $\Xi:=\Ga(h_0)\setminus\seqeqnat{\{\neg F,F_y(\suc y)\}}$)
\begin{align*}
*(h) 
 &= \emptyset \cup
     \Big( {\Ga({\ind^{y,n,i}_{F}h_0 })}
        \setminus\seqeqnat{\{F_y(\ul{n+2^i})\}} \Big) \\
 &\quad \cup
     \Big( {\Ga({\ind^{y,n+2^i,i}_{F}h_0 })}
        \setminus\seqeqnat{\{\neg F_y(\ul{n+2^i})\}} \Big) \\
 &=
     \Big( \big( \{\neg F_y(\ul n), F_y(\ul{n+2^i})\} 
              \cup \Xi \big)
        \setminus\seqeqnat{\{F_y(\ul{n+2^i})\}} \Big)  \\
 &\quad \cup
     \Big( \big( \{\neg F_y(\ul{n+2^i}),F_y(\ul{n+2^{i+1}}) \} 
              \cup \Xi \big)
        \setminus\seqeqnat{\{\neg F_y(\ul{n+2^i})\}} \Big) \\
 &\subseteq
     \{\neg F_y(\ul n),F_y(\ul{n+2^{i+1}})\} 
              \cup \Xi \\
 &= {\Ga(h)}
\end{align*}

\medskip

\noi
\textbf{Case 9.}
$\IS=\Cut_C$:
$\tp(h)=\Cut_C$ and 
$h[\iota]=h_\iota$ for $\iota<2$,
hence
\begin{align*}
*(h) 
 &=  \emptyset
  \cup ({\Ga({h_0})}\setminus\seqeqnat{\{C\}})
  \cup ({\Ga({h_1})}\setminus\seqeqnat{\{\neg C\}}) \\
 &={\Ga(h)}
\end{align*}

\end{proof}

\begin{observation}
The following relations and functions are polynomial time computable:
the finitary proof system $\bastar$,
the set of $\bastar$-quasi derivations and the functions
$h\mapsto\Ga(h)$, $h\mapsto\hgt(h)$, and $h\mapsto\dsz{h}$ denoting 
the endsequent,
the height and the size for a $\bastar$-quasi derivation $h$;
the bounding term $t\mapsto\bd(t)$ for terms $t$ occurring in $\ForNBA$
and the relations $\bd(h)\le m$ and $\ibd(h)\le m$ on
$\DerNBA\times\NN$;
the set $\DerNBA$ and the functions $h\mapsto\tp(h)$, 
$h,i\mapsto h[i]$, $h\mapsto\Ga(h)$,
$m,h\mapsto\ord_m(h)$ and $h\mapsto|h|$.
\end{observation}
\begin{proof}
For bounding terms we use our assumption that a fixed (finite)
number of function symbols and term depth is only allowed, which
implies that terms can only denote a fixed finite number of 
different polynomial time computable functions.
That $\bd(h)\le m$ is polynomial time computable is clear
as the computation of $\bd(h)$ computes a monotone increasing 
sequence of values by successively applying one of the finitely many
polynomial time computable functions, and once the bound $m$ is
exceeded during this process we can already output \emph{NO}.
\end{proof}

As the function $\bd(h)$ in general may not be polynomially bounded,
we cannot conclude in general
that $\ord(h)$ is polynomial time computable.
However, the function $m,h\mapsto\ord_{\min(|\ibd(h)|,m)}(h)$ is
polynomial time computable and will be sufficient in our applications.

\end{section}

\begin{section}{Computational content of proofs}

Let us start by describing the idea for computing witnesses using 
proof trees.
Assume we have a BA proof of an existential formula 
$(\exists y)\vhi(y)$ and we want to compute a $k$ such that 
$\vhi(k)$ is true  
-- in case we are interested in definable functions, such a situation
is obtained from a proof of $(\forall x)(\exists y)\vhi(x,y)$ by
inverting the universal quantifier to some $n\in\NN$.
Assume further, we have applied some proof theoretical transformations to
obtain a $\bainf$ derivation $\derd$ of $(\exists y)\vhi(y)$
with $\Ccrk(\derd)\le\Crk(\vhi)$ for some set of formulae $\cC$ (the
choice of $\cC$ depends on the level of definability we are interested in).
Then we can define a path through $\derd$, represented by sub-derivations
\[  \derd=\derd_0,\derd_1,\derd_2,\dots  \]
with
\begin{itemize}
\item
$\derd_{\ell+1}=\derd_\ell(i)$ for some $i\in|\last(d_\ell)|$
\item
$\Ga(\derd_\ell)= (\exists y)\vhi(y), \Ga_\ell$  where all formulae 
$A\in\Ga_\ell$ are false and satisfy $\Crk(A)\le\Crk(\vhi)$.
\end{itemize}

As $\derd$ is well-founded, such a path must be finite, i.e.\ ends with 
some $\derd_\ell$ say.
In this situation we must have that 
$\last(\derd_\ell)=\tbv^k_{(\exists y)\vhi(y)}$ 
and that $\vhi(k)$ is true.
Hence we can output $k$.

Such a path can be viewed as the canonical path to the following 
local search problem:
Let $F$ be a set of possible solutions, 
which is a subset of $\bainf$ containing only those $d'$ which satisfy
that $\Ga(d')\subseteq \{(\exists y)\vhi(y)\}\cup\Ga'$ 
where all formulae $A\in\Ga'$ are false and satisfy $\Crk(A)\le\Crk(\vhi)$.
Furthermore, assume $d\in F$ and 
that $F$ is closed under the following neighbourhood
function $N\colon\bainf\to\bainf$ which is defined by case distinction
on the shape of $\last(d')$ for $d'\in F$:

\begin{itemize}
\item
$\last(\derd)=\Ax_A$ cannot occur as 
all atomic formulae in $\Ga(\derd')$ are false.
\item
$\last(\derd)=\tbw_{A_0\land A_1}$, then $A_0\land A_1$ must be 
false, hence some of $A_0,A_1$ must be false.
Let $N(\derd):= \derd(0)$ if $A_0$ is false, and 
$\derd(1)$ otherwise.
\item
$\last(\derd)=\tbw_{A_0\lor A_1}$, then $A_0\lor A_1$ must be 
false, hence both $A_0,A_1$ must be false.
Let $N(\derd):= \derd(0)$.
\item
$\last(\derd)=\tbw_{(\forall x)A(x)}$.
As $(\forall x)A(x)$ is false there is some $i$ 
such that $A(i)$ is false.
Let $N(\derd):=\derd(i)$.
\item
$\last(\derd)=\tbv^k_{(\exists x)A(x)}$.
If $(\exists x)A(x)$ is different from $(\exists y)\vhi(y)$
then $(\exists x)A(x)$ must be false; let $N(d):=d(0)$.
Otherwise, let $N(d)=d(0)$ in case $\vhi(k)$ is false, and
$N(d)=d$ in case it is true (in which case we found a true solution 
to the original search problem).
\item
$\last(\derd_\ell)=\Cut_{C}$.
If $C$ is false let $N(\derd):= \derd(0)$, otherwise let 
$N(\derd):= \derd(1)$.
\end{itemize}

The idea in the following will be to use proof notations from $\DerNBA$
to denote this search problem.
This way we will obtain characterisations of the definable functions of 
Bounded Arithmetic theories.

The level of proof theoretic reduction will be adjusted in such a 
way that occurring formulae which have to be decided fall exactly 
in the computational class under consideration.
So our main concern in order for this strategy to be meaningful 
is to find feasible upper bounds for the length of such reduction 
sequences and for the complexity of derivation notations occurring in them.

\subsection{Complexity notions for \bastar}

In order to handle the complexity of \bastar proof notations occurring in the
set of possible solutions, we need some notions describing key 
complexity properties of them which we will provide first.

Although $\tp(A)=\tbw$ for any $A$ starting with a $\forall$,
and thus we can denote infinitely many direct sub-formulae by $A[n]$ for all
$n\in\NN$, only finitely many carry non-trivial information, because
all quantifiers (and in particular this outermost $\forall$) are bounded.
The next definition makes this formal by assigning first to each closed 
formula in $\ForNBA$, then to each inference symbol in $\bainf$, and finally
to each proof notation in $\ceDerNBA$, its range.

\begin{DEF}
Let $A$ be a formula in \ForNBA.
We define \emph{the range of $A$,} denoted $\rng(A)$, by
\[
\rng(A):=\begin{cases}
0 &  \text{if } A \text{ a literal} \enspace, \\
2 &  \text{if } A=B\land C \text{ or } A=B\lor C \enspace, \\
t^\NN+1 &  \text{if } A=(\forall x\le t)B \text{ or } A=(\exists x\le t)B 
\enspace.
\end{cases}
\]

Let $\IS$ be an inference symbol of \bainf.
We define \emph{the range of $\IS$,} denoted $\rng(\IS)$, by
\[
\rng(\IS):=\begin{cases}
0 &  \text{if } \IS=\Ax_A \enspace, \\
1 &  \text{if } \IS=\tbv^k_C \text{ or } \IS=\Rep \enspace, \\
\rng(C) &  \text{if } \IS=\tbw_C \enspace, \\
2 &  \text{if } \IS=\Cut_C \enspace.
\end{cases}
\]

For $h\in\ceDerNBA$ we define
\[
\rng(h) := \rng(\tp(h)) \enspace.
\]
\end{DEF}

\begin{DEF}
We extend the definition of bounding terms $\bd(h)$ and $\ibd(h)$ 
from $\DerNBA$ to $\ceDerNBA$ in the following way
by induction on the build-up of $h\in\ceDerNBA$:
\begin{itemize}
\item
If $h\in\DerNBA$ then the definition of $\bd(h)$ and $\ibd(h)$ 
are inherited from the definition of $\bd$ resp.\ $\ibd(h)$ on \DerNBA.
\item
If $h=\I^k_C h_0$ then
\begin{align*}
\bd(h)&:=
\begin{cases}
\bd(h_0) & \text{if } k<\rng(C) \enspace, \\
0 & \text{otherwise} \enspace.
\end{cases} \\
\ibd(h)&:= \ibd(h_0)
\end{align*}

\item
$\bd(\R_C h_0 h_1):=\max\{\bd(h_0),\bd(h_1)\}$,
$\ibd(\R_C h_0 h_1):=\max\{\ibd(h_0),\ibd(h_1)\}$.

\item
$\bd(\E h_0) := \bd(h_0)$,
$\ibd(\E h_0) := \ibd(h_0)$.
\end{itemize}
\end{DEF}

\begin{lem}\label{lem:bd}
Let $h\in\ceDerNBA$.
\begin{enumerate}
\item
If $j<\rng(h)$ then $\bd(h[j])\le\bd(h)$ and $\ibd(h[j])\le\ibd(h)$.
\item
If $\tp(h)=\tbv^k_C$ then $k\le\bd(h)$.
\end{enumerate}
\end{lem}

\begin{proof}[Proof by induction on the build-up of $h$.]
\end{proof}

\begin{DEF}
For $h\in\bastar\cup\ceDerNBA$ we define \emph{the set of decorations of $h$,
$\deco(h)\in\finpot(\BFOR)$,} by induction on the build-up of $h$.
Let $h=\IS h_0\dots h_{n-1}$.
We define
\[
\deco(h):=\deco(\IS)\Big(\bigcup_{i<n}\deco(h_i)\Big)
\]
where
\[
\deco(\IS)(S):=
\begin{cases}
S\cup\De(\IS)\cup\{F\}
  & \text{if } \IS=\ind^{y,t}_{F} \text{ or } \IS=\ind^{y,a,i}_{F} \enspace, \\
S\cup\De(\IS) & \text{otherwise} \enspace.
\end{cases}
\]
\end{DEF}

\begin{observation}
We have $\Ga(h)\subseteq\deco(h)$.
\end{observation}

\begin{DEF}
Let $\compDerNBA$ be the set of all $h\in\ceDerNBA$ which have 
the property that all occurrences of $\I^k_C$ in $h$ satisfy $k<\rng(C)$.
\end{DEF}

\begin{DEF}
Let $\Phi$ be a finite set of formulae in $\BFOR$, 
and let $K\in\NN$ be a size parameter.
With $\Phi_K$ we denote the set of formulae which result 
from formulae in $\Phi$ by substituting free variables by constants from
$\setof{c_i}{0\le i\le K}$.
\end{DEF}

\begin{lem}\label{lem:deco}
Let $h\in\compDerNBA$ and $\Phi\in\finpot(\BFOR)$ such that
$\deco(h)\subseteq\Phi$, and $\Phi$ is closed under negation 
and taking sub-formulae.
Let $j,K\in\NN$ and $y$ be a variable.
\begin{enumerate}
\item
If $j\le K$ and $C\in\Phi$, then $C[j]\in\Phi_K$.
\item
If $j\le K$ then $\deco(h(\ul j/y))\subseteq\Phi_K$.
\item
$\De(\tp(h))\subseteq\deco(h)_{\bd(h)}$ 
(subscript $\bd(h)$ needed e.g.\ for $\ind^{y,n,i+1}_F$).
\item
If $j<\rng(h)$ 
then $\deco(h[j])\subseteq\Phi_{\bd(h)}$.
\end{enumerate}
\end{lem}

\begin{proof}
For 4., consider the case that $h=\R_C h_0 h_1$, $\tp(h_1)=\tbv^k_{\neg C}$
and $j=0$, i.e.\ $h[0]=\I^k_C h_0$.
By 3.\ we have $\neg C\in\Phi_{\bd(h_1)}$, hence $C\in\Phi_{\bd(h)}$.
Also $k\le\bd(h_1)$ by Lemma~\ref{lem:bd}, 2.
Hence, $C[k]\in\Phi_{\bd(h)}$ by 1.
Now we compute
\[
  \deco(h[0]) = \{C[k]\}\cup\deco(h_0) \subseteq
    \Phi_{\bd(h)}\cup\Phi = \Phi_{\bd(h)} \enspace.
\]
\end{proof}

\begin{lem}
For $h\in\ceDerNBA$ we have that the cardinality of $\Ga(h)$ is
bounded above by $2\cdot\dsz{h}$.
\end{lem}
\begin{proof}
Let the cardinality of a set $S$ be denoted by $\card(S)$.
We observe that $\card(\De(\IS))\le2$ for any $\IS\in\bainf$.
Thus we can compute for $h=\IS h_0\dots h_{n-1}\in\ceDerNBA$
\[
\card(\Ga(h)) \le \card(\De(\IS)) +\sum_{i<n}\card(\Ga(h_i))
 \le 2+\sum_{i<n}2\cdot\dsz{h_i} = 2\cdot\dsz{h}
\enspace. \]
\end{proof}

\subsection{Search problems defined by proof notations}

We identify the notation system $\DerNBA$ for $\bainf$
with the abstract system of proof notations associated with
it according to Observation~\ref{obs:abstraction-of-notation-system}.
For $s\in\NN$ a size parameter we define
\[
  \DerNBA^s := \setof{h\in\DerNBA}{\sz{h}\le s} \enspace.
\]
Then $\DerNBA^s$ is an $s$-bounded, abstract system of proof 
notations, because we observe that
$h\in\DerNBA$ and $h\to h'$ implies $\sz{h'}\le\sz{h}$.

Remember that $\abstr{h}$ for $h\in\ceDerNBA$ denotes the abstraction
of $h$ which allows us to view $\ceDerNBA$ as a subsystem of $\Ecl\DerNBA$
(see Definition~\ref{def:abstraction} and 
Observation~\ref{obs:abstraction}).

\begin{DEF}
For $h\in\ceDerNBA$ we define $\szf h(s) := \szf{\abstr h}(s)$.
\end{DEF}

Theorem~\ref{th:main} now reads as follows:
\begin{cor}\label{cor:main}
If $h\in\ceDerNBA^s$ and $h\to h'$, then
$\szf h(s)\geq \szf{h'}(s)$.
\end{cor}

\begin{DEF}\label{def:local-search}
We define a local search problem $L$ parameterised by
\begin{itemize}
\item
a finite set of bounded formulae $\Phi\subset\BFOR$,
\item
a \emph{``complexity class'' $\cC$} given as a polynomial time computable 
set of $\lba$-formulae 
(usually $\cC=\sib i$ for some $i$),
\item
a \emph{size parameter $s\in\NN$,}
\item
an \emph{initial value function $h_\cdot\colon \NN\to\compDerNBA^s$,}
where $h_a$ is presented in the form $\E\dots\E h(\ul a/x)$ for
some $\bastar$-derivation $h$,
\item
a formula $(\exists y)\vhi(x,y)\in\Phi$ with $\neg\vhi\in\cC$,
\end{itemize}
such that, for $a\in\NN$,
\begin{itemize}
\item
$\Ga(h_a)=\{(\exists y)\vhi(\ul a,y)\}$,
\item
$\Ccrk(h_a)\le1$,
\item
$\ord(h_a)=2^{|a|^{O(1)}}$,
\item
$\szf{h_a}(s) = |a|^{O(1)}$,
\item
$\deco(h_a)\subseteq\Phi_a$,
\end{itemize}
in the following way:

\begin{itemize}
\item
The set of \emph{possible solutions $F(a)\in\finpot(\compDerNBA^s)$}
is given as the set of those $h\in\compDerNBA^s$ which satisfy:
\begin{enumerate}
\item[i)]
$\Ga(h)\subseteq\{(\exists y)\vhi(\ul a,y)\}\cup\De$ for some
$\De\subseteq\cC\cup\neg\cC$ such that all $A\in\De$ are closed and false,
\item[ii)]
$\Ccrk(h)\le1$,
\item[iii)]
$\ord(h)\le\ord(h_a)$,
\item[iv)]
$\szf h(s)\le\szf{h_a}(s)$,
\item[v)]
$\bd(h)\le\bd(h_a)$ and $\ibd(h)\le\ibd(h_a)$,
\item[vi)]
$\deco(h)\subseteq\Phi_{\bd(h_a)}$;
\end{enumerate}

\item
The \emph{initial value function} is given by $i(a):=h_a$;

\item
the \emph{cost function} is defined as $c(a,h):=\ord(h)$; and

\item
the \emph{neighbourhood function} is given by
\[
N(a,h):=
\begin{cases}
h[j] & \text{if } \tp(h)=\tbw_C, j<\rng(C)
        \text{ and } C[j]\text{ false}\enspace,\\
h[0] & \text{if } \tp(h)=\tbv^i_C \text{ and } C\neq(\exists y)\vhi(\ul a,y) \\
     & \text{ or } \tp(h)=\tbv^i_{(\exists y)\vhi(\ul a,y)}
        \text{ and } \vhi(\ul a,\ul i)\text{ false} \enspace, \\
h[0] & \text{if } \tp(h)=\Cut_C \text{ and } C\text{ false} \enspace, \\
h[1] & \text{if } \tp(h)=\Cut_C \text{ and } C\text{ true} \enspace, \\
h[0] & \text{if } \tp(h)=\Rep \enspace, \\
h & \text{otherwise} \enspace.
\end{cases}
\]
\end{itemize}
(Observe that the just defined neighbourhood function is a multi-function
due to case $\tbw_C$.)
\end{DEF}
\begin{proof}
First observe that the initial value is indeed a possible solution,
$i(a)=h_a\in F(a)$.

Let $h\in F(a)$, $h':=N(a,h)$. 
Then we show
\begin{enumerate}
\item
$h\neq h'$ implies $h\to h'$ and $\ord(h')<\ord(h)$,
\item
$h'\in F(a)$.
\end{enumerate}
For $h=h'$ the assertions are obvious.
So let us assume $h\neq h'$.
Then $h'=h[j]$ for some $j<\rng(h)$ by construction.
Hence, the first claim is obvious.

For the second claim, we consider i)--vi) of the definition of $h'\in F(a)$:
ii) is clear; 
iii) is obvious; 
for iv) observe that $h\to h'$, thus
$\szf{h'}(s)\le\szf{h}(s)$ by Corollary~\ref{cor:main};
for v) observe that $j<\rng(h)$ implies $\bd(h')\le\bd(h)$ and
$\ibd(h')\le\ibd(h)$ by Lemma~\ref{lem:bd};
for vi) observe that $j<\rng(h)$ implies 
$\deco(h')\subseteq(\Phi_{\bd(h_a)})_{\bd(h)}=\Phi_{\bd(h_a)}$
by Lemma~\ref{lem:deco}, 2., because $\bd(h)\le\bd(h_a)$.
And finally for i) we first observe that 
the first condition that 
$\Ga(h)\setminus\{(\exists y)\vhi(\ul a,y)\}$ is a subset of $\cC\cup\neg\cC$
consisting only of closed formulae,
is satisfied, as $\Ccrk(h)\le1$.
For the second condition of i) let $\IS:=\tp(h)$.
We have by Proposition~\ref{prop:cons-faithfull} that
\[
  \Ga(h[j])\subseteq \seqeqnat{\Big(\Ga(h)\cup\De_j(\IS)\Big)}
\]
thus it is enough to show that $\tbv\De_j(\IS)$ is false.
\begin{itemize}
\item
$\IS=\tbw_C$: 
$\De_j(\IS)=\{C[j]\}$ and $C[j]$ false by construction.
\item
$\IS=\tbv^i_C$: then $j=0$.
If $C\neq(\exists y)\vhi(\ul a,y)$, then $\De_0(\IS)=\{C[i]\}$.
Now $C$ is false by i) of $h\in F(a)$, hence $C[i]$ must be false as well.
Otherwise, $\De_0(\IS)=\{\vhi(\ul a,\ul i)\}$, and $\vhi(\ul a,\ul i)$ false
by construction.
\item
$\IS=\Cut_C$: If $j=0$, then $\De_0(\IS)=\{C\}$ and $C$ false by construction.
Otherwise, $j=1$, then $\De_1(\IS)=\{\neg C\}$ and $\neg C$ false
by construction.
\item
$\IS=\Rep$:
then $j=0$ and $\De_0(\IS)=\emptyset$ and nothing is to show.
\end{itemize}
\end{proof}

\begin{prop}[Complexity of $L$]
$F\in\Pt^\cC$, $i,c\in\FP$, and $N\in\FP^\cC[\wit,1]$.
\end{prop}
\begin{proof}
First observe that the functions $a\mapsto i(a)=h_a$,
$a\mapsto\bd(h_a)$, $a\mapsto\ibd(h_a)$, $a\mapsto\ord(h_a)$,
$a\mapsto\szf{h_a}$, and $a\mapsto\deco(h_a)$ are
polynomial time computable.

Furthermore, the relations $\compDerNBA^s$,
$\Ccrk(h)\le1$, $\bd(h)\le m$, $\ibd(h)\le m$ and
$\deco(h)\subseteq\Phi_m$ are polynomial time computable,
and once $\ibd(h)\le m$ is established we also can
compute $\ord(h)\le m'$ and then $\ord(h)$ in polynomial time.
Hence $c\in\FP$

Also, the functions $\tp(h)$ and $h[i]$ are polynomial time computable
on $\ceDerNBA$, which shows $N\in\FP^\cC[\wit,1]$.

For $F\in\Pt^\cC$ observe that 
$\Ga(h)\subseteq\deco(h)\subseteq\Phi_{\bd(h_a)}$,
hence condition $h\in F(a)$, i), is a property in $\Pt^\cC$.
\end{proof}

\begin{prop}[Properties of $L$]\label{prop:properties-L}
\begin{enumerate}

\item
$N(a,h)=h$ implies $\tp(h)=\tbv^i_{(\exists y)\vhi(\ul a,y)}$
with $\vhi(\ul a,\ul i)$ true.
Thus, the local search problem $L$ defines a multi-function
by mapping $a$ to $i$ (this is called the computed multi-function).

\item
The search problem $L$ in general defines a search problem in 
$\text{PLS}^\cC$,
assuming that we turn the neighbourhood (multi-)function into a real
function, which can easily be achieved by using an intermediate
$\text{PLS}^\cC$ search problem which looks for the smallest witness
for the case $\tp(h)=\tbw_C$.
Then $N\in\FP^\cC$.

\item
Assume $\ord(h_a)=|a|^{O(1)}$.
Then the canonical path through $L$, which starts at $h_a$ and leads to a 
local minimum, is of polynomial length with terms of polynomial size, 
thus the computed multi-function is in $\FP^\cC[\wit,\ord(h_a)]$.
\end{enumerate}
\qed
\end{prop}

\subsection{\sib i-definable multi-functions in $\rS^{i-1}_2$}

Let $i\ge2$ and assume that $\rS^{i-1}_2\vdash(\forall x)(\exists y)\vhi(x,y)$
with $(\exists y)\vhi(x,y)\in\sib i$, $\vhi\in\pib{i-1}$.
By partial cut-elimination we obtain some \bastar-derivation $h$ such that
\begin{itemize}
\item
$\fv(h)\subseteq\{x\}$,
\item
$\Ga(h)=\{(\exists y)\vhi(x,y)\}$,
\item
$\ocrk{\sib{i-1}}(h)\le1$, and
\item
$\ord(h(\ul a/x)) = O(||a||)$.
\end{itemize}
We define a search problem by stating its parameters:
\begin{itemize}
\item
$\Phi:=\deco(h)$ is a finite set of formulae in $\BFOR$,
\item
as the ``complexity class'' we take $\cC:=\sib{i-1}$,
\item
for the size parameter we choose $s:=\sz{h}$,
\item
the initial value function is given by $h_a:=h(\ul a/x)$,
\item
the formula is as given, $(\exists y)\vhi(x,y)$.
\end{itemize}
This defines a local search problem according to 
Definition~\ref{def:local-search}, because
\begin{itemize}
\item
$\Ga(h_a)=\Ga(h(\ul a/x))=\Ga(h)(\ul a/x)=\{(\exists y)\vhi(\ul a,y)\}$,
\item
as $h\in\DerNBA^s$ we have $h(\ul a/x)\in\DerNBA^s$, hence
$\szf{h_a}(s) = s = O(1)$
\item
$\deco(h_a)\subseteq\Phi_a$ by Lemma~\ref{lem:deco}, 1.
\end{itemize}

As $\ord(h_a)=O(||a||)$, Proposition~\ref{prop:properties-L}, 3., shows
that the computed multi-function of this search problem is in
$\FP^{\sib{i-1}}[\wit,O(\log n)]$,
which coincides with the description given by \krajicek \cite{krajicek:93}.

\subsection{\sib i-definable functions in $\rS^i_2$}

Let $i>0$ and assume that $\rS^i_2\vdash(\forall x)(\exists y)\vhi(x,y)$
with $(\exists y)\vhi(x,y)\in\sib i$, $\vhi\in\pib{i-1}$.
By partial cut-elimination we obtain some \bastar-derivation $h$ such that
\begin{itemize}
\item
$\fv(h)\subseteq\{x\}$,
\item
$\Ga(h)=\{(\exists y)\vhi(x,y)\}$,
\item
$\ocrk{\sib{i-1}}(h)\le2$, and
\item
$\ord(h(\ul a/x)) = O(||a||)$.
\end{itemize}
We define a search problem by stating its parameters:
\begin{itemize}
\item
$\Phi:=\deco(h)$ is a finite set of formulae in $\BFOR$,
\item
as the ``complexity class'' we take $\cC:=\sib{i-1}$,
\item
for the size parameter we choose $s:=\sz{h}$,
\item
the initial value function is given by $h_a:=\E h(\ul a/x)$,
\item
the formula is as given, $(\exists y)\vhi(x,y)$.
\end{itemize}
This defines a local search problem according to 
Definition~\ref{def:local-search}, because
\begin{itemize}
\item
$\Ga(h_a)=\{(\exists y)\vhi(\ul a,y)\}$,
\item
$\ocrk{\sib{i-1}}(h_a)\le1$,
\item
$\ord(h_a)=2^{\ord(h(\ul a/x))}-1= 2^{O(||a||)}=|a|^{O(1)}$,
\item
as $h(\ul a/x)\in\DerNBA^s$ we have
\begin{align*}
\szf{h_a}(s) &= \szf{\E h(\ul a/x)}(s) \\
  &= \ord(h(\ul a/x))\cdot(\szf{h(\ul a/x)}(s)+2) \\
  &= O(||a||)\cdot(s+2) = O(||a||)
\end{align*}
\item
$\deco(h_a)\subseteq\Phi_a$.
\end{itemize}

As $\ord(h_a)=|a|^{O(1)}$, Proposition~\ref{prop:properties-L}, 3., shows
that the computed multi-function of this search problem is in
$\FP^{\sib{i-1}}[\wit,n^{O(1)}]=\FP^{\sib{i-1}}[\wit]$.

But this immediately implies that the \sib i-definable functions 
of $\rS^i_2$ are in $\FP^{\sib{i-1}}$, because a witness query to
$(\exists z<t)\psi(u,z)$ can be replaced by $|t|$ many usual (non-witness)
queries to $\chi(a,b,u) = (\exists z<t)(a\le z<b\land\psi(u,z))$
using a divide and conquer strategy.
This characterisation coincides with the one given by Buss \cite{buss:86}.

\subsection{\sib i-definable multi-functions in $\rS^{i+1}_2$}

Let $i>0$ and assume that $\rS^{i+1}_2\vdash(\forall x)(\exists y)\vhi(x,y)$
with $(\exists y)\vhi(x,y)\in\sib i$, $\vhi\in\pib{i-1}$.
By partial cut-elimination we obtain some \bastar-derivation $h$ such that
\begin{itemize}
\item
$\fv(h)\subseteq\{x\}$,
\item
$\Ga(h)=\{(\exists y)\vhi(x,y)\}$,
\item
$\ocrk{\sib{i-1}}(h)\le3$, and
\item
$\ord(h(\ul a/x)) = O(||a||)$.
\end{itemize}
We define a search problem by stating its parameters:
\begin{itemize}
\item
$\Phi:=\deco(h)$ is a finite set of formulae in $\BFOR$,
\item
as the ``complexity class'' we take $\cC:=\sib{i-1}$,
\item
for the size parameter we choose $s:=\sz{h}$,
\item
the initial value function is given by $h_a:=\E \E h(\ul a/x)$,
\item
the formula is as given, $(\exists y)\vhi(x,y)$.
\end{itemize}
This defines a local search problem according to 
Definition~\ref{def:local-search}, because
\begin{itemize}
\item
$\Ga(h_a)=\{(\exists y)\vhi(\ul a,y)\}$,
\item
$\ocrk{\sib{i-1}}(h_a)\le1$,
\item
$\ord(h_a)=2^{\ord(\E h(\ul a/x))}-1= 2^{|a|^{O(1)}}$,
\item
as $h(\ul a/x)\in\DerNBA^s$ we have
\begin{align*}
\szf{h_a}(s) &= \szf{\E \E h(\ul a/x)}(s) \\
  &= \ord(\E h(\ul a/x))\cdot(\szf{\E h(\ul a/x)}(s)+2) \\
  &= |a|^{O(1)}\cdot(O(||a||)+2) = |a|^{O(1)}
\end{align*}
\item
$\deco(h_a)\subseteq\Phi_a$.
\end{itemize}

By Proposition~\ref{prop:properties-L}, 2., this defines a search problem
in $\text{PLS}^{\sib{i-1}}$.
This coincides with the description given by 
Buss and \krajicek \cite{buss:krajicek:94}.

\subsection{\sib{i+1}-definable multi-functions in $\sib{i+j}\dilind{2+j}$}

Let $i\ge1$, $j\ge0$, and assume that 
$\sib{i+j}\dilind{2+j}\vdash(\forall x)(\exists y)\vhi(x,y)$
with $(\exists y)\vhi(x,y)\in\sib{i+1}$, $\vhi\in\pib{i}$.
By partial cut-elimination we obtain some \bastar-derivation $h$ such that
\begin{itemize}
\item
$\fv(h)\subseteq\{x\}$,
\item
$\Ga(h)=\{(\exists y)\vhi(x,y)\}$,
\item
$\ocrk{\sib{i}}(h)\le j+1$, and
\item
$\ord(h(\ul a/x)) = O(|a|_{3+j})$.
\end{itemize}
We define a search problem by stating its parameters:
\begin{itemize}
\item
$\Phi:=\deco(h)$ is a finite set of formulae in $\BFOR$,
\item
as the ``complexity class'' we take $\cC:=\sib{i}$,
\item
for the size parameter we choose $s:=\sz{h}$,
\item
the initial value function is given by 
$h_a:=\underbrace{\E\dots \E}_{j\text{ times}}h(\ul a/x)$,
\item
the formula is as given, $(\exists y)\vhi(x,y)$.
\end{itemize}
This defines a local search problem according to 
Definition~\ref{def:local-search}, because
\begin{itemize}
\item
$\Ga(h_a)=\{(\exists y)\vhi(\ul a,y)\}$,
\item
$\ocrk{\sib{i}}(h_a)\le1$,
\item
$\ord(h_a)\le 2_j({\ord(h(\ul a/x))})= 2_j(\OO{|a|_{3+j}})$,
\item
as $h(\ul a/x)\in\DerNBA^s$ we have
\begin{align*}
\szf{h_a}(s) &= \szf{\underbrace{\E\dots\E}_{j\times} h(\ul a/x)}(s) \\
  &= \ord(\underbrace{\E\dots\E}_{(j-1)\times} h(\ul a/x))\cdot
(\szf{(\underbrace{\E\dots\E}_{(j-1)\times} h(\ul a/x)}(s)+2) \\
  &= 2_{j-1}(\OO{|a|_{3+j}})\cdot
(\szf{(\underbrace{\E\dots\E}_{(j-1)\times} h(\ul a/x)}(s)+2) \\
  &= \dots = \OO{|a|}
\end{align*}
\item
$\deco(h_a)\subseteq\Phi_a$ by Lemma~\ref{lem:deco}, 1.
\end{itemize}

As $\ord(h_a)=O(||a||)$, Proposition~\ref{prop:properties-L}, 3., shows
that the computed multi-function of this search problem is in
$\FP^{\sib{i}}[\wit,2_j(\OO{\log^{2+j} n})]$, which coincides with the
description given by Pollett~\cite{pollett:99}.

\end{section}

\section*{Acknowledgements}

The authors gratefully acknowledge support by the
Engineering and Physical Sciences Research Council (EPSRC) 
under grant number EP/D03809X/1.

\end{document}